\documentclass[aps,twocolumn,prx,showpacs,notitlepage,superscriptaddress,longbibliography]{revtex4-2}
\usepackage{graphicx}
\usepackage{xcolor}
\usepackage{enumerate}
\usepackage{amsmath, amssymb}
\usepackage{stmaryrd}
\usepackage{multirow}
\usepackage{float}
\usepackage{dsfont}
\usepackage{amssymb}
\usepackage{placeins}
\usepackage[pdftex,colorlinks=true,linkcolor=blue,citecolor=blue]{hyperref}

\usepackage{algorithm2e}
\RestyleAlgo{ruled}

\usepackage{amssymb}

\usepackage{soul}
\setstcolor{red}

\usepackage[normalem]{ulem}
\newcommand{\stkout}[1]{\ifmmode\text{\sout{\ensuremath{#1}}}\else\sout{#1}\fi}

\DeclareMathAlphabet\mathbfcal{OMS}{cmsy}{b}{n}
\newcommand*{\mybox}[2][5cm]{%
  \makebox[#1][c]{#2}}

\begin{document}
\title{
Exploring the quantum capacity of a Gaussian random displacement channel using Gottesman-Kitaev-Preskill codes and maximum likelihood decoding
}

\author{Mao Lin}
\affiliation{Amazon Braket, Seattle, WA, USA, 98170, USA}%
\author{Kyungjoo Noh}
\email[Corresponding author: ]{nkyungjo@amazon.com}
\affiliation{AWS Center for Quantum Computing, Pasadena, CA 91125, USA}

\date{\today}

\begin{abstract}

Determining the quantum capacity of a noisy quantum channel is an important problem in the field of quantum communication theory. In this work, we consider the Gaussian random displacement channel $\mathcal{N}_{\sigma}$, a type of bosonic Gaussian channels relevant in various bosonic quantum information processing systems. In particular, we attempt to make progress on the problem of determining the quantum capacity of a Gaussian random displacement channel by analyzing the error-correction performance of several families of multi-mode Gottesman-Kitaev-Preskill (GKP) codes. In doing so we analyze the surface-square GKP codes using an efficient and exact maximum likelihood decoder (MLD) up to a large code distance of $d=39$. We find that the error threshold of the surface-square GKP code is remarkably close to $\sigma=1/\sqrt{e}\simeq 0.6065$ at which the best-known lower bound of the quantum capacity of $\mathcal{N}_{\sigma}$ vanishes. We also analyze the performance of color-hexagonal GKP codes up to a code distance of $d=13$ using a tensor-network decoder serving as an approximate MLD. By focusing on multi-mode GKP codes that encode just one logical qubit over multiple bosonic modes, we show that GKP codes can achieve non-zero quantum state transmission rates for a Gaussian random displacement channel $\mathcal{N}_{\sigma}$ at larger values of $\sigma$ than previously demonstrated. Thus our work reduces the gap between the quantum communication theoretic bounds and the performance of explicit bosonic quantum error-correcting codes in regards to the quantum capacity of a Gaussian random displacement channel. 
\end{abstract}

\maketitle

\section{Introduction}

Quantum communication \cite{bennett1992communication,Bennett1993_teleporting} and quantum computing \cite{shor1994algorithms,lloyd1996universal,nielsen2002quantum} are two intricately related fields within quantum information science, aiming to leverage the principles of quantum mechanics to revolutionize data transmission and information processing. One common challenge in these areas is the need to protect the encoded quantum information from noise in quantum systems. Quantum error correction (QEC) thus plays a pivotal role in both fields by providing a mean to systematically suppress the effects of noise. Owing to this close relationship, the concepts and methodologies developed in one area can often be applied to another. For example, noisy processes in quantum computers can be modeled as noisy quantum channels for transmitting quantum information. This analogy is particularly fruitful when it is complemented by a general framework of quantum communication theory. In particular, the latter allows one to study maximum achievable data transmission rates of a noisy quantum channel through various entropic quantities \cite{schumacher1996quantum,lloyd1997capacity,barnum1998information,Devetak2005,wilde2013quantum}, providing fundamental bounds on the capability of a QEC code for correcting errors in the noisy quantum channel. 

Bosonic Gaussian channels \cite{weedbrook2012gaussian} have been extensively studied in the field of quantum communication theory \cite{holevo2001evaluating,gottesman2001encoding,Harrington2001_achievable,harrington2004analysis,Caruso2006,Wolf2007,Holevo2007,Pirandola2017,Sharma2018,Rosati2018,Noh2019,Noh2020_enhanced,Fanizza2021,Mele2022,Mele2024} due to their relevance to realistic optical and microwave quantum communication channels. Moreover dominant errors in bosonic modes in various quantum computing platforms (e.g., 2D resonators \cite{Putterman2024_hardware} and 3D cavities \cite{Paik2011,Reagor2013,Milul2023} in superconducting systems, motional modes in trapped-ion systems \cite{fluhmann2019encoding,Matsos2024}, and light modes in optical systems \cite{Madsen2022}) are well described by bosonic Gaussian channels. Thus, numerous bosonic QEC codes \cite{Cochrane1999Macroscopically,Jeong2002Efficient,gottesman2001encoding,Leghtas2013_hardware,Michael2016,albert2018performance} have been proposed to correct for errors in these bosonic Gaussian channels. 

In this work, we explore the interface between quantum communication theory and quantum error correction through bosonic Gaussian channels and bosonic QEC. In particular, we consider the problem of determining the quantum capacity \cite{schumacher1996quantum,lloyd1997capacity,barnum1998information,Devetak2005} $Q(\mathcal{N}_{\sigma})$ of a Gaussian random displacement channel $\mathcal{N}_{\sigma}$, i.e., its maximum achievable quantum state transmission rate per channel use under an optimal QEC strategy. 
The quantum capacities of some special cases of bosonic Gaussian channels, such as the pure-loss and pure-amplification channels, are well understood analytically \cite{holevo2001evaluating,Wolf2007,Pirandola2017,Sharma2018} thanks to their degradability or anti-degradability property \cite{Caruso2006}. However generic bosonic Gaussian channels including the Gaussian random displacement channel do not possess this special property. Thus only lower \cite{holevo2001evaluating,Noh2020_enhanced} and upper \cite{holevo2001evaluating,Pirandola2017,Sharma2018,Rosati2018,Noh2019} bounds on their quantum capacity are known. 

Here we attempt to make progress on this long-standing open problem by investigating the performance of multi-mode Gottesman-Kitaev-Preskill (GKP) codes \cite{Harrington2001_achievable,harrington2004analysis,fukui2017analog,fukui2018high, fukui2019high,vuillot2019quantum,noh2020fault,hanggli2020enhanced,royer2022encoding,conrad2022gottesman,conrad2023good,lin2023closest} against errors caused by a Gaussian random displacement channel $\mathcal{N}_{\sigma}$. Specifically, we aim to find a tighter lower bound on the quantum capacity of the Gaussian random displacement channel by searching for explicit QEC schemes which achieve higher quantum state transmission rates than the best-known lower bound on $Q(\mathcal{N}_{\sigma})$. Our search is based on multi-mode GKP codes and their maximum likelihood decoding. 

\section{Gaussian random displacement channel and scaled self-dual GKP codes}
In the Heisenberg picture, a Gaussian random displacement channel $\mathcal{N}_{\sigma}$ transforms the position and momentum operators $\hat{q}$ and $\hat{p}$ of a bosonic mode into $\hat{q}+\xi_{q}$ and $\hat{p}+\xi_{p}$, respectively. Here, $\xi_{q}$ and $\xi_{p}$ are independent classical noise variables following a Gaussian distribution with vanishing mean and a variance of $\sigma^{2}$. Note that a Gaussian random displacement channel is also referred to as an additive Gaussian noise channel in the literature. 

The seminar results in the field of quantum communication theory established that the quantum capacity $Q(\mathcal{N})$ of a quantum channel $\mathcal{N}$ equals its regularized coherent information, i.e., 
\begin{align}
    Q(\mathcal{N}) = \lim_{N\rightarrow \infty} \frac{1}{N} \max_{\hat{\rho}} I_{c}(\mathcal{N}^{\otimes N}, \hat{\rho}) .  \label{eq:regularized_coherent_information}
\end{align}
Here, $I_{c}(\mathcal{N},\hat{\sigma}) \equiv S(\mathcal{N}(\hat{\sigma})) - S(\mathcal{N}^{c}(\hat{\sigma}))$ is the coherent information of the channel $\mathcal{N}$ with respect to an input state $\hat{\sigma}$, $S(\hat{\sigma}) \equiv -\text{Tr}[\hat{\sigma}\log_{2}\hat{\sigma}]$ is the quantum von Neumann entropy, and $\mathcal{N}^{c}$ is the complementary channel of $\mathcal{N}$ \cite{schumacher1996quantum,lloyd1997capacity,barnum1998information,Devetak2005}. However in practice, evaluating this entropic quantity is difficult due to the maximization over all possible input states $\hat{\rho}$ as well as the need to consider an asymptotic limit of $N\rightarrow \infty$. 

For the specific case of a Gaussian random displacement channel, the quantum capacity is bounded to be 
\begin{align}
     \max[ \log_{2}\Big{(} \frac{1}{e\sigma^{2}} \Big{)},  0] \le Q(\mathcal{N}_{\sigma})   \le  \max[\log_{2}\Big{(} \frac{1 - \sigma^{2}}{\sigma^{2}} \Big{)},  0] , \label{eq:Gaussian_random_displacement_channel_capacity_bounds}
\end{align}
where the lower bound is obtained by computing its one-shot coherent information (i.e., a similar expression as in Eq.~\eqref{eq:regularized_coherent_information} but with $N=1$,  see App.~\ref{sec:derivation_lower_bound} for a derivation) \cite{holevo2001evaluating} and the upper bound is derived using a data-processing inequality \cite{Noh2019}. From the lower bound and the definition of the quantum capacity, it follows that there must exist a QEC scheme that achieves a quantum state transmission rate of $\log_{2}( \frac{1}{e\sigma^{2}})$ for the Gaussian random displacement channel $\mathcal{N}_{\sigma}$ with $\sigma \le 1/\sqrt{e}\simeq 0.6065$. Earlier works showed that scaled self-dual GKP codes can achieve the rate of $\log_{2}( \lfloor \frac{1}{e\sigma^{2}} \rfloor )$ \cite{Harrington2001_achievable,harrington2004analysis}. Notably, the floor function here is due to the fact that scaled self-dual GKP codes are restricted to cases where every single one of the $N$ normal modes encodes the same number (a positive integer $\lambda$) of logical states, thereby encoding $\lambda^{N}$ logical states or $N\log_{2}\lambda$ logical qubits overall. Thus when $\frac{1}{e\sigma^{2}}$ is given by a positive integer (e.g., $\frac{1}{e\sigma^{2}} = 2 \leftrightarrow \sigma = 1/\sqrt{2e} \simeq 0.4289$), the scaled self-dual GKP codes achieve the lower bound in Eq.~\eqref{eq:Gaussian_random_displacement_channel_capacity_bounds} exactly. However in all other cases, the achievable rate of the scaled self-dual GKP codes falls short of the lower bound of $Q(\mathcal{N}_{\sigma})$. Thus in what follows, we aim to reduce the gap between the lower bound of $Q(\mathcal{N}_{\sigma})$ and the achievable rate of GKP codes. In particular we focus on the large $\sigma$ regime (e.g., $0.5 \le \sigma \le 1/\sqrt{e} \simeq 0.6065$) and also discuss the prospect of achieving a non-zero rate in the range $1/\sqrt{e} \le \sigma < 1/\sqrt{2}$, thereby improving on the best-known lower bound, since this possibility is not ruled out by the upper bound in Eq.~\eqref{eq:Gaussian_random_displacement_channel_capacity_bounds}.     

\section{Achievable rates of concatenated multi-mode GKP codes}
Here we investigate the performance of various concatenated multi-mode GKP codes such as the surface-square GKP codes, the $[[5,1,3]]$-hexagonal GKP code, and the color-hexagonal GKP codes. Importantly, we focus on codes that encode only one logical qubit over the entire $N$ modes for reasons to be made clear below. Moreover we implement a maximum likelihood decoder (MLD) \cite{vuillot2019quantum,conrad2022gottesman} either exactly \cite{bravyi2014efficient} or approximately using a tensor-network decoder \cite{bravyi2014efficient,chubb2021general} such that we can probe the ultimate limits on the error-correction capability of these concatenated GKP codes against displacement errors.
We review MLD in the context of GKP codes in App.~\ref{sec:Maximum likelihood decoder for GKP codes}.

\begin{figure}[b!]
\centering
\includegraphics[width=\linewidth]{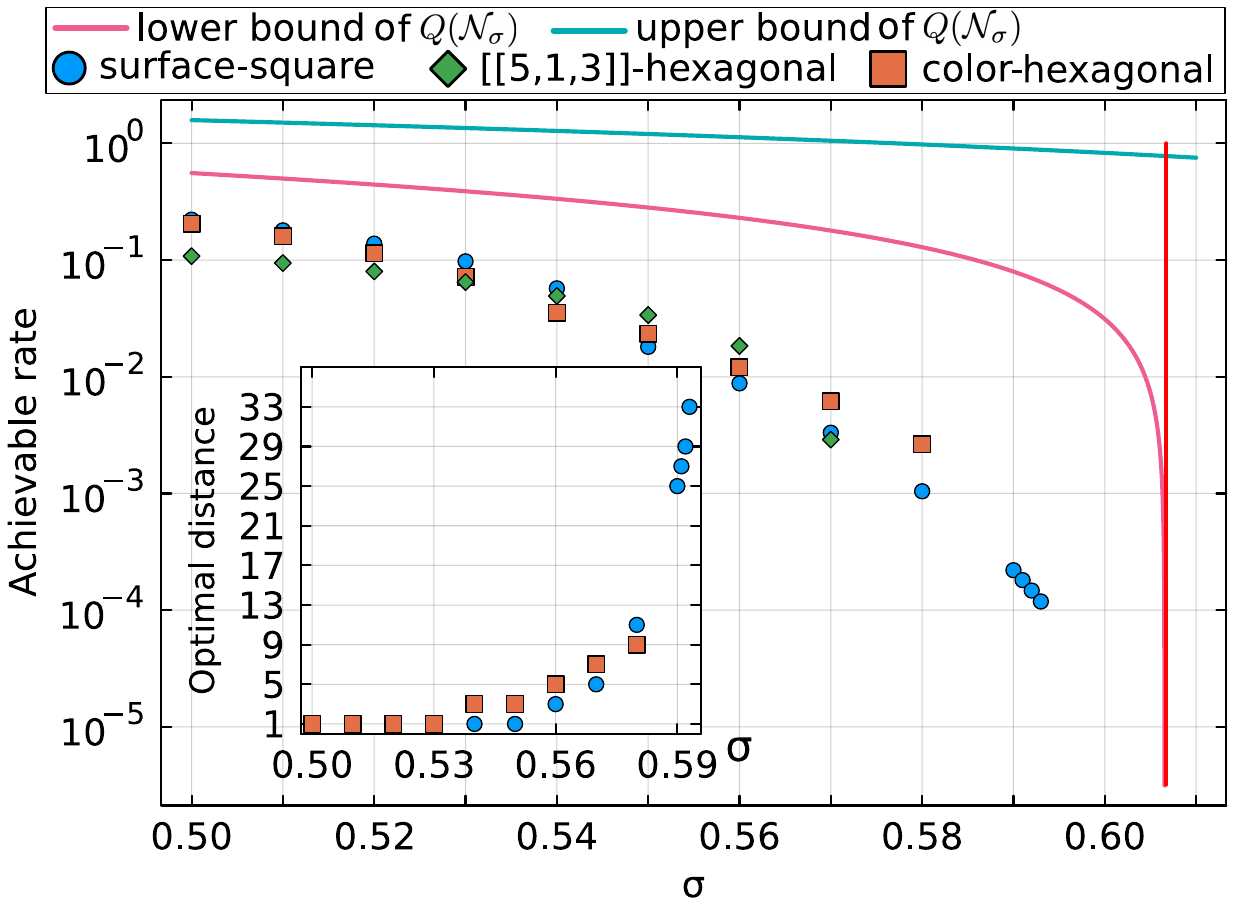}
 \caption{Achievable rates of the surface-square, $[[5,1,3]]$-hexagonal, color-hexagonal GKP codes for the Gaussian random displacement channel $\mathcal{N}_{\sigma}$ as a function of $\sigma$. The value of $\sigma$ is scanned from $0.50$ to $0.59$ in steps of $0.01$, and from $0.590$ to $0.600$ in steps of $0.001$. Each data point is obtained with $10^6-10^7$ Monte Carlo samples in the decoding operation. For reference, the lower and upper bounds of the quantum capacity of Gaussian random displacement channel are also shown. The red vertical line indicates $\sigma = 1/\sqrt{e}\simeq 0.6065$ at which the best-known lower bound of $Q(\mathcal{N}_{\sigma})$ vanishes. }
 \label{fig:achievable_rate_and_capacity_bounds}
\end{figure}

In Fig.~\ref{fig:achievable_rate_and_capacity_bounds}, we show the achievable rates of the surface-square, $[[5,1,3]]$-hexagonal, color-hexagonal GKP codes as a function of the noise standard deviation $\sigma$. Since all these instances of multi-mode GKP codes encode one logical qubit over $N$ modes, we compute the achievable rate of these codes using the hashing bound \cite{bennett1996mixed,wilde2013quantum}, i.e.,
\begin{align}
    \label{eq:achievable_rate_GKP_as_Pauli_divided_N_maintext}
    R = \frac{1}{N}I_{c}\left(\mathcal{P}_{\vec{p}}, \frac{1}{2}\hat{I}\right)=\frac{1}{N}\left(1 + \sum_{\mu=I,X,Y,Z}p_\mu\log_2 p_\mu\right).
\end{align}
Here, $\hat{I}/2$ is the maximally-mixed qubit state and $\vec{p}=(p_I,p_X,p_Y,p_Z)$ are parameters of the single-qubit Pauli error channel
$\mathcal{P}_{\vec{p}}$. The channel is obtained by applying an $N$-mode GKP encoding, $N$ independent copies of Gaussian random displacement channels, and a decoding operation. The parenthesis in Eq.~\eqref{eq:achievable_rate_GKP_as_Pauli_divided_N_maintext} is the hashing bound of Pauli channels (see App.~\ref{sec: More details on the hashing bound of Pauli channels} for the derivation), which is divided by $N$ to reflect the fact that $N$ copies of $\mathcal{N}_{\sigma}$ are consumed to produce each $\mathcal{P}_{\vec{p}}$. 

For the surface-square GKP codes, we use an exact MLD based on an idea similar to the one given in Ref.~\cite{bravyi2014efficient} adapted to GKP codes. For the $[[5,1,3]]$-hexagonal GKP code, given its small code size, we perform brute-force calculations of the coset probabilities of all four logical single-qubit Pauli operators to perform MLD exactly (see App.~\ref{sec:More details on the brute force MLD for the 513-hexagonal GKP code} for more details). Lastly for the color-hexagonal GKP codes, we perform MLD approximately by using a variant of the tensor-network decoder in Ref.~\cite{chubb2021general,SweepContractor_v0_1_7} adapted to GKP codes. The approximation is made when truncating the bond dimensions to a maximum allowed cutoff (64 in our work) during tensor-network contractions. See App.~\ref{sec: More details on the tensor-network decoder for the color-GKP codes} for more details on the decoding procedures we use and Ref.~\cite{latticealgorithms} for the source code.  

Note that for the families of surface-square and color-hexagonal GKP codes, we sweep the distance of an outer code, hence also sweeping the total number of modes $N$ used ($N=d^{2}$ for the surface code and $N=(3d^{2}+1)/4$ for the color code). Then in the main plot, we report the highest achievable rate optimized over the code distance. The optimal code distance maximizing the achievable rate is shown in the inset. For $\sigma \lesssim 0.53$, the optimal code distance is given by $d=1$, indicating that the single-mode square and hexagonal GKP codes achieve higher quantum state transmission rates than larger instances (i.e., $d \ge 3$) of the surface-square and color-hexagonal GKP codes, respectively. This is because with $\sigma \lesssim 0.53$, the fidelities of these single-mode GKP codes are sufficiently high and thus the protection offered by an outer code is not worthwhile given the extra overhead in terms of a large number of modes $N$. 

In contrast as the noise standard deviation $\sigma$ increases, multi-mode GKP codes begin to outperform single-mode GKP codes thanks to the increased protection against displacement errors provided by an outer error-correcting code. For example around $\sigma\simeq 0.55$, the $[[5,1,3]]$-hexagonal GKP code achieves the highest quantum state transmission rate among all the instances of multi-mode GKP codes we considered. With $\sigma\ge 0.57$, the surface-square and color-hexagonal GKP codes achieve higher rates. Notably as $\sigma$ increases, the optimal code distance that maximizes the achievable rate increases for both the surface-square and color-hexagonal GKP code families. 

The results in Fig.~\ref{fig:achievable_rate_and_capacity_bounds} thus clearly show the particular relevance of one-logical-qubit-into-many-oscillators codes when searching for GKP codes that might achieve (or even surpass) the best-known lower bound of $Q(\mathcal{N}_{\sigma})$ in the large $\sigma$ regime, i.e., $\sigma \sim 1/\sqrt{e}\simeq 0.6065$. In particular, our results provide a non-trivial demonstration that GKP codes can achieve non-zero quantum state transmission rates for a Gaussian random displacement channel $\mathcal{N}_{\sigma}$ at larger values of $\sigma$ than previously demonstrated \cite{gottesman2001encoding,Harrington2001_achievable,harrington2004analysis} if a large number of modes (e.g., $N=1521$ for the $d=39$ surface-square GKP code) is used to encode just one logical qubit. 

The largest multi-mode GKP code we could analyze numerically is the $d=39$ surface-square GKP code which achieves a non-zero quantum state transmission rate at a value of $\sigma$ as large as $\sigma=0.598$ (see App.~\ref{sec: More data for the coherent information for the concatenated-GKP codes}). However, higher performance may be achieved with an even larger number of modes or with another code family outperforming the surface-square GKP code family. In the rest of the paper, we provide additional numerical data to shed further light on the prospect of achieving (or surpassing) the lower bound of $Q(\mathcal{N}_{\sigma})$ with multi-mode GKP codes.   

\begin{figure}[b!]
\centering
\includegraphics[width=\linewidth]{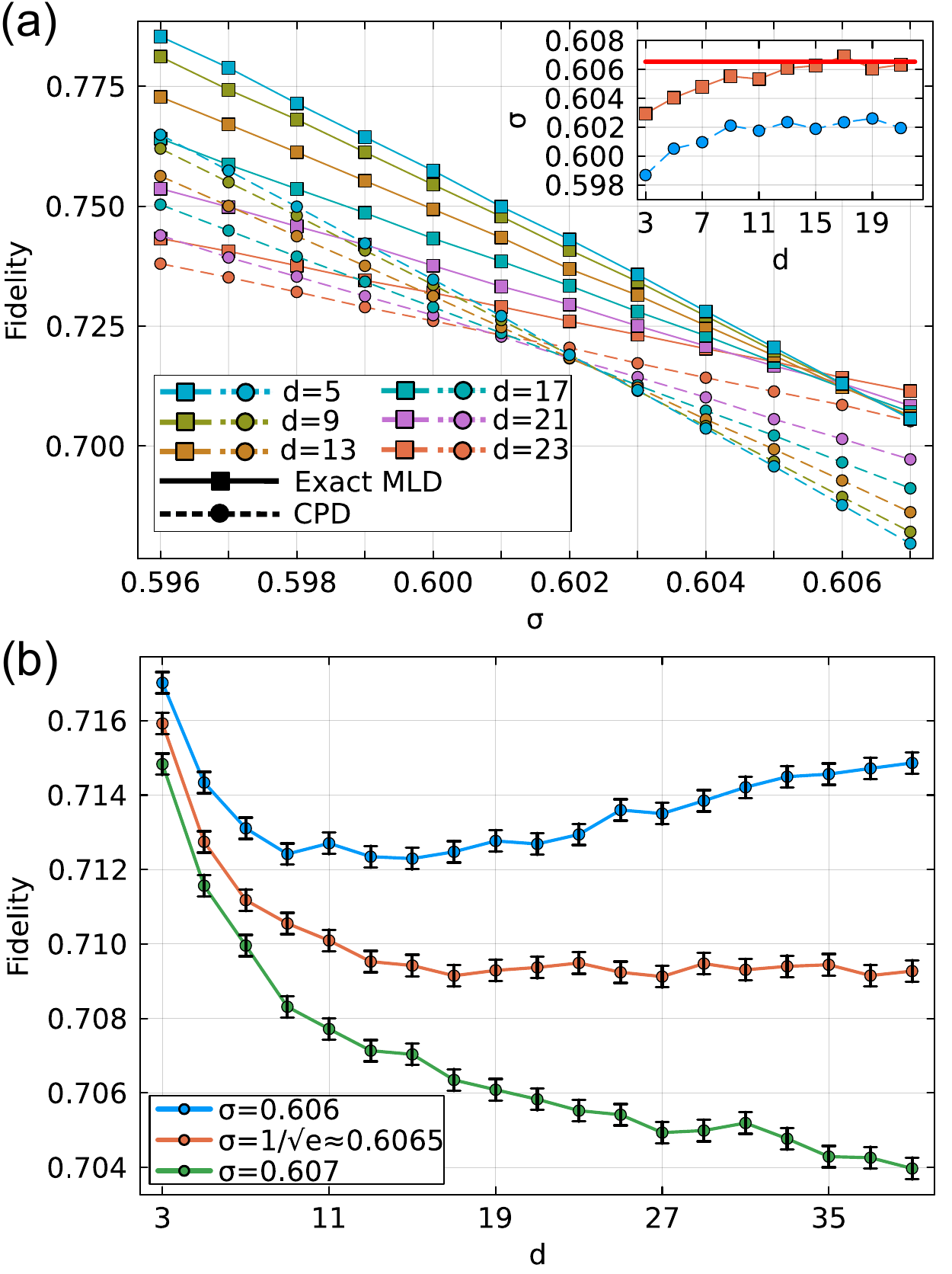}
 \caption{(a) The fidelity of the surface-square GKP code with distances between $d=5$ and $d=23$. The dashed and solid lines correspond to the fidelity obtained from the closest-point decoder (CPD) and the exact maximum likelihood decoder (MLD). 
 The inset shows the value of $\sigma$ at which the fidelities with code distances $d$ and $d+2$ cross as a function of $d$ for both CPD and exact MLD. These crossing points are used to infer the error threshold of the surface-square GKP code against $\mathcal{N}_{\sigma}$. The red solid line corresponds to $\sigma \equiv 1/\sqrt{e} \simeq 0.6065$. 
 (b) The fidelity of the surface-square GKP code as a function of the code distance $d$ at $\sigma=0.606, 1/\sqrt{e}, 0.607$. Each data point is obtained with $10^7$ Monte Carlo samples.
}
 \label{fig:surface_square_performance}
\end{figure}

\section{Exact MLD for the surface-square GKP codes}
We begin by illustrating the importance of using an exact MLD as opposed to a suboptimal decoder when exploring the quantum capacity of a Gaussian random displacement channel with GKP codes. In Fig.~\ref{fig:surface_square_performance}(a), we compare the performance of the surface-square GKP code with two different decoders, i.e., the closest-point decoder (CPD; dashed line) \cite{lin2023closest} and the exact MLD (solid line) based on Ref.~\cite{bravyi2014efficient}. 
More details on the exact MLD can be found in App.~\ref{app: Review the BSV decoders}.
Since CPD converges to MLD only in the $\sigma\rightarrow 0$ limit, MLD outperforms CPD in the large $\sigma$ regime where $Q(\mathcal{N}_{\sigma})$ nearly vanishes. Interestingly when the exact MLD is used, the error threshold of the surface-square GKP code is very close to $\sigma = 1/\sqrt{e} \simeq 0.6065$ at which the best-known lower bound of $Q(\mathcal{N}_{\sigma})$ vanishes.     

To get further insights into the asymptotic behavior of the surface-square GKP code in the $d\rightarrow \infty$ limit, we show in Fig.~\ref{fig:surface_square_performance}(b) the fidelity of the surface GKP code as a function of the code distance up to $d=39$ for three different values of $\sigma$, i.e., $\sigma \in [0.606, 1/\sqrt{e}\simeq 0.6065, 0.607]$. Remarkably, at exactly $\sigma = 1/\sqrt{e}\simeq 0.6065$, the fidelity plateaus within the statistical variation (caused by a finite number of Monte Carlo samples) as the code distance increases. Moreover at a slightly smaller (or larger) value of $\sigma$, the fidelity increases (or decreases) as the code distance $d$ increases. These numerical observations suggest that for any $\sigma< 1/\sqrt{e}$, the surface-square GKP code may be below the error threshold of $\mathcal{N}_{\sigma}$ and may consequently achieve a non-zero quantum state transmission rate with a sufficiently large code distance $d$. Note that our method of determining the threshold of the surface-square GKP code is more rigorous compared to the ones used in related prior works \cite{fukui2017analog, fukui2018high, hanggli2020enhanced, zhang2022concatenation}, which studied several families of GKP codes against $\mathcal{N}_{\sigma}$ near $\sigma = 1/\sqrt{e}\simeq 0.6065$ (See App.~\ref{sec:Comparison to prior works} for further discussions).

\section{Approximate MLD for color-hexagonal GKP codes using a tensor-network decoder}
\begin{figure}[t!]
\centering
\includegraphics[width=\linewidth]{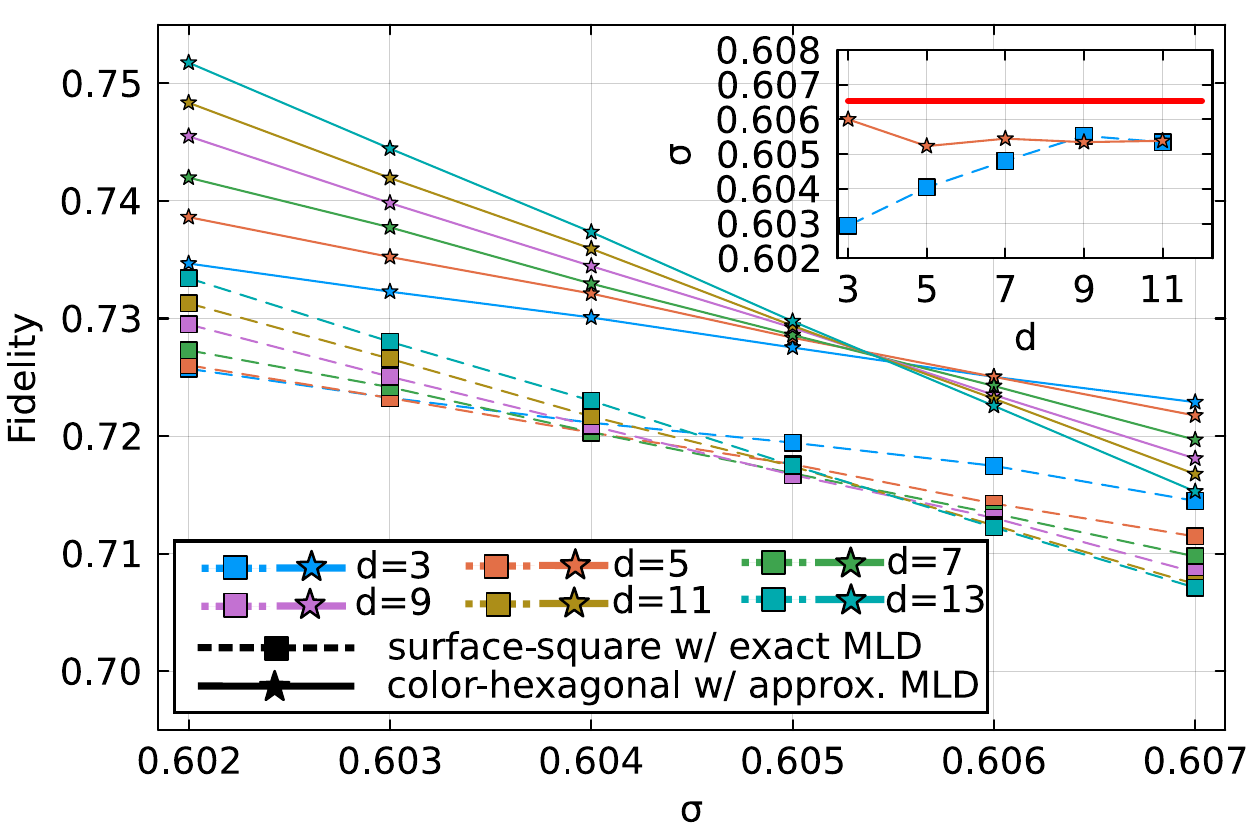}
 \caption{Comparison of the fidelities of the color-hexagonal GKP code decoded with a tensor-network decoder (solid line) and the surface-square GKP code decoded with the exact MLD (dash line) for distances between $d=3$ and $d=13$. The inset shows crossing points as a function of $d$ similarly as in Fig.~\ref{fig:surface_square_performance}(a). The red solid line corresponds to $\sigma = 1/\sqrt{e}\simeq 0.6065$.  
}
 \label{fig:color_hex_vs_surf_sq}
\end{figure}

As shown in Fig.~\ref{fig:achievable_rate_and_capacity_bounds}, the color-hexagonal GKP code family sometimes outperforms the surface-square GKP code family in terms of the achievable rate. We thus provide additional numerical data on the color-hexagonal GKP codes. In Fig.~\ref{fig:color_hex_vs_surf_sq}, we show the fidelities of the color-hexagonal GKP code using an approximate MLD and the surface-square GKP code using the exact MLD. The approximate MLD for the color-hexagonal GKP code is implemented by using a tensor-network decoder \cite{chubb2021general,SweepContractor_v0_1_7} with a maximum bond dimension of $64$ during tensor-network contractions. We observe that despite the suboptimal, approximate MLD, the color-hexagonal GKP code family outperforms the surface-square GKP code family in the regime of $0.602\le \sigma \le 0.607$ at the same respective code distance $d$. When combined with the promising asymptotic trend of the surface-square GKP code family shown in Fig.~\ref{fig:surface_square_performance}(b), the numerical results in Fig.~\ref{fig:color_hex_vs_surf_sq} suggest that the asymptotic performance (in the $d\rightarrow \infty$ limit) of the color-hexagonal GKP code may have an important implication on the quantum capacity of the Gaussian random displacement channel. It is thus an interesting open question as to whether the error threshold of the color-hexagonal GKP code family can exceed $\sigma = 1/\sqrt{e}$ or not (especially when an exact MLD is used) and correspondingly whether a non-zero quantum state transmission rate can be acheived for a Gaussian random displacement channel $\mathcal{N}_{\sigma}$ with $\sigma\ge 1/\sqrt{e}$. A positive result in this direction will establish an improved lower bound of $Q(\mathcal{N}_{\sigma})$.  

\section{Conclusion and outlook}
In this work, we have considered an open problem of determining the quantum capacity of a Gaussian random displacement channel (i.e., $Q(\mathcal{N}_{\sigma})$). In particular we have aimed to make progress on this problem by analyzing concrete instances of multi-mode GKP codes using either an exact or approximate maximum likelihood decoder. Our results clearly illustrate the particular relevance of GKP codes which encode just one logical qubit over multiple bosonic modes when studying a Gaussian random displacement channel $\mathcal{N}_{\sigma}$ in a high noise regime (e.g., $\sigma\sim 1/\sqrt{e}$). 
 
Our numerical results also show that the error threshold of the surface-square GKP code is remarkably close to $\sigma = 1/\sqrt{e} \simeq 0.6065$ at which the lower bound of $Q(\mathcal{N}_{\sigma})$ vanishes. Moreover we show that the color-hexagonal GKP code generally outperforms the surface-square GKP code in a similar regime of $\sigma$. Thus our results motivate further research into the asymptotic behaviors of the surface-square and color-hexagonal GKP codes in the $d\rightarrow \infty$ limit, especially in relation to the quantum capacity of a Gaussian random displacement channel.  

Note that the maximum likelihood decoding of a quantum error-correcting code can often be mapped to a statistical mechanical model \cite{Dennis2002,Chubb2021_statistical} and this connection has been explored in the context of toric-GKP codes \cite{vuillot2019quantum} as well. An interesting avenue for future research is to extend this relationship further and attempt to connect this mapping also to the problem of quantum communication theory since our work reveals a close connection between the maximum likelihood decoding of the surface-square and color-hexagonal GKP codes and the problem of Gaussian quantum channel capacity. 

Lastly, an earlier work \cite{lin2023closest} provided several instances of numerically-optimized multi-mode GKP codes that outperform the known concatenated GKP codes, even including the surface-square GKP code and the color-hexagonal GKP code, with the same number of modes. The performances of these codes have only been analyzed through a suboptimal closest-point decoder. Thus, continued search of new multi-mode GKP codes beyond concatenation and their performance analysis using maximum-likelihood decoder can also be fruitful future research directions, potentially having relevance to the problem of Gaussian quantum channel capacity as well.

\section{Acknowledgements}
We would like to acknowledge the AWS EC2 resources
which were used for part of the numerical simulations performed in
this work. ML would like to thank Péter Kómár and Eric Kessler for their supports of the project. We also would like to thank the Day 1 Science Mentorship (D1SM) program led by Connor Stewart at Amazon through which this project was initiated.

\appendix

\section{More data for the coherent information for the concatenated-GKP codes}
\label{sec: More data for the coherent information for the concatenated-GKP codes}

\begin{figure}[!ht]
\centering
\includegraphics[width=\linewidth]{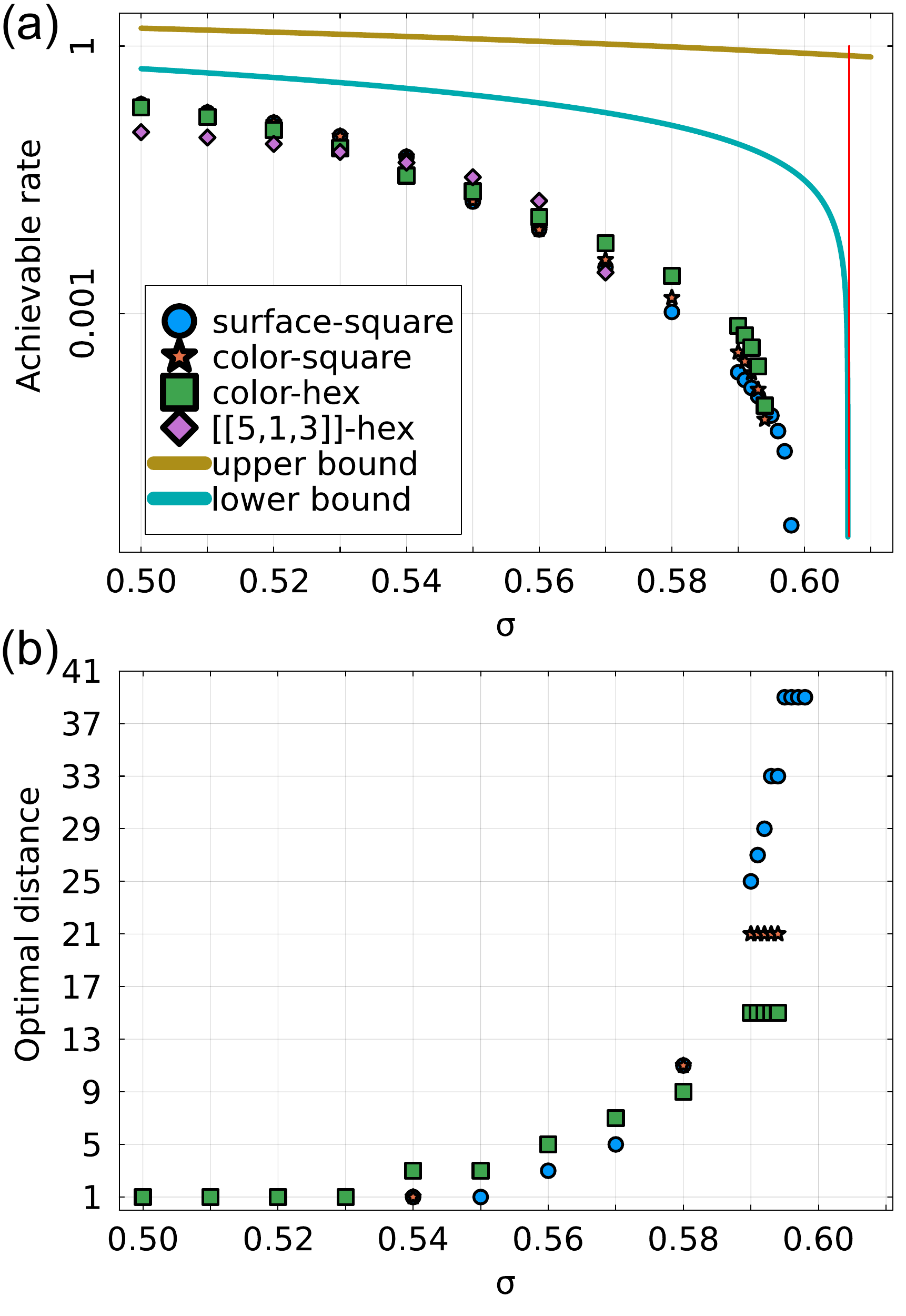}
 \caption{(a) An expanded version of Fig.~\ref{fig:achievable_rate_and_capacity_bounds} of the main text with color-square GKP codes included. Moreover, additional data points in the large $\sigma$ regime are included for comprehensiveness despite the fact that an optimal outer code distance could not be found due to the saturation caused by a finite maximum code distance analyzed in numerics. 
 (b) An expanded version of the inset in Fig.~\ref{fig:achievable_rate_and_capacity_bounds}, showing the optimal code distances that maximizes the achievable rates.} 
 \label{fig:achievable_rate_and_capacity_bounds_raw_data}
\end{figure}

In Fig.~\ref{fig:achievable_rate_and_capacity_bounds_raw_data}, we display more data for the coherent information for the three concatenated-GKP codes discussed in the main text, in addition to the color-square GKP code. For surface-square GKP code, we have collected data up to $d=39$, and found that the code could have positive achievable rate up to $\sigma=0.598$. The corresponding optimal distances are shown in Fig.~\ref{fig:achievable_rate_and_capacity_bounds_raw_data}(b), and we found that the code has optimal distance $d=39$ at $0.595\leq\sigma\leq0.598$. This is clearly an artifact of the fact that we have only simulated the code up to $d=39$. As a result, we have omitted the last 5 data points for surface-square GKP code in Fig.~\ref{fig:achievable_rate_and_capacity_bounds_raw_data} and report the rest of the data in Fig.~\ref{fig:achievable_rate_and_capacity_bounds} in the main text. 

Similarly, we have collected data up to $d=21$ and $d=15$ for the color-square and color-hexagonal GKP codes respectively, and found that they could have positive achievable rate up to $\sigma=0.594$, much larger than $0.58$ shown in Fig.~\ref{fig:achievable_rate_and_capacity_bounds}. 
We have omitted color-square code in Fig.~\ref{fig:achievable_rate_and_capacity_bounds} for clarity because it does not exhibit better achievable rate than other GKP codes for the range of $\sigma$ considered.
Further, we have consistently removed the last 5 data points from the color-hexagonal code in Fig.~\ref{fig:achievable_rate_and_capacity_bounds}, because of the similar plataeu observed in the plot of optimal distances.

The ``finite-size'' effects are due to the limited compute resource and we emphasize that it is important to omit the data points affected by these effects when trying to interpret our numerical data. Otherwise, one would may make inaccurate conclusions about the achievable rate of the GKP codes. For example from Fig.~\ref{fig:achievable_rate_and_capacity_bounds_raw_data}(a), it appears that the achievable rates of both color-square and color-hexagonal codes drop much quickly than that of the surface-square code for $\sigma>0.59$. However, this is an artifact of having only considered finite distances up to $d=21$ for the color-square GKP code and $d=15$ for the color-hexagonal GKP code. 

\begin{figure}[!ht]
\centering
\includegraphics[width=\linewidth]{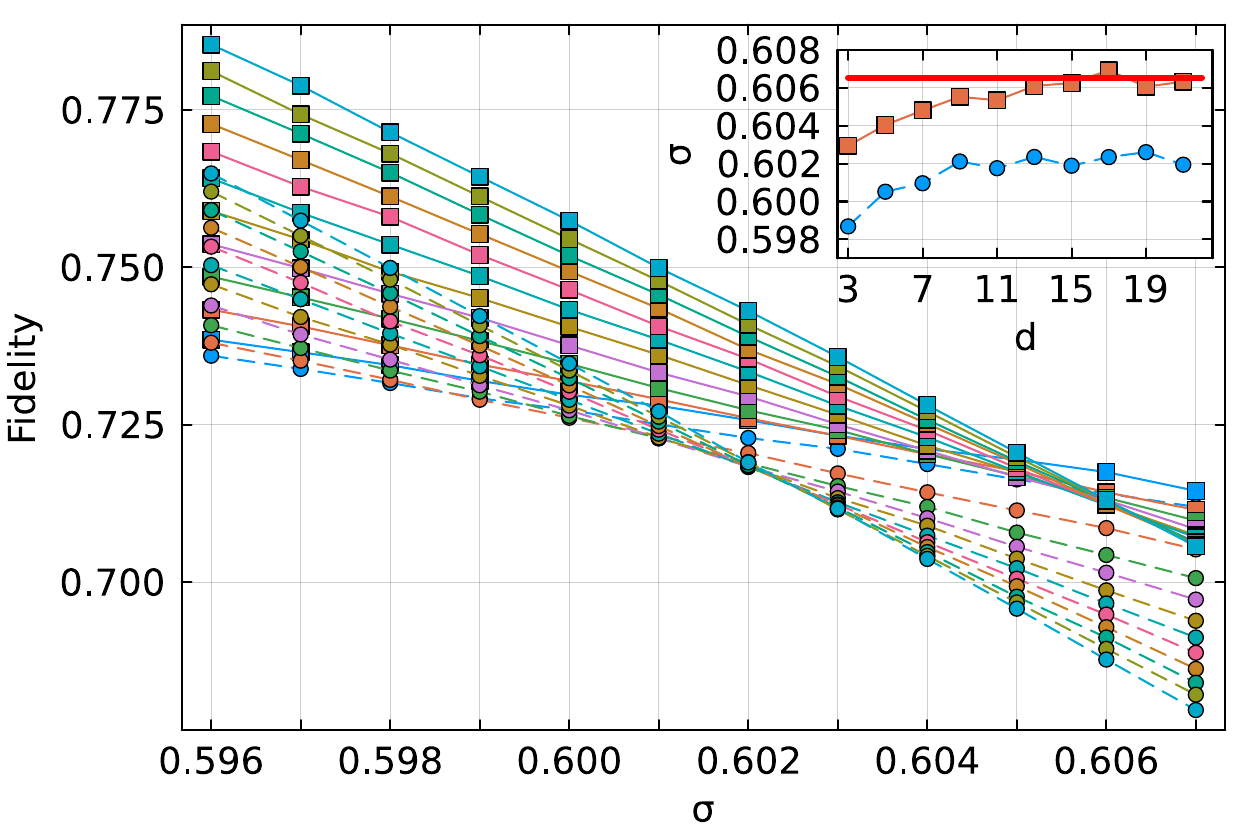}
 \caption{An expanded version of Fig.~\ref{fig:surface_square_performance}(a) with all the raw data presented.
}
 \label{fig:6}
\end{figure}

In Fig.~\ref{fig:6}, we provide an expanded version of Fig.~\ref{fig:surface_square_performance}(a) with all the data from $d=3$ to $d=23$ displayed.

\section{Maximum likelihood decoding for GKP codes}
\label{sec:Maximum likelihood decoder for GKP codes}

In this section, we review MLD for GKP codes following Ref.~\cite{conrad2022gottesman}.
Let $\Lambda$ denote the lattice corresponding to the GKP code considered, where the lattice points $\boldsymbol{u}\in\sqrt{2\pi}\Lambda$ correspond to the stabilizers of the GKP code. The prefactor of $\sqrt{2\pi}$ ensures the stabilizers commute with each other. 
For the purpose of decoding, we will need to consider the symplectic dual lattice $\Lambda^\perp$ where the quotient 
\begin{align}
\label{eq:def_logical_operator}
    \sqrt{2\pi}(\Lambda^\perp/\Lambda)\equiv\left\{\boldsymbol{l}^\perp\right\}    
\end{align}
consists of the lattice points in $\sqrt{2\pi}\Lambda^\perp$ (and not in $\sqrt{2\pi}\Lambda$) that correspond to the logical operators. 
Throughout this work, we assume the quadrature variables are subject to independent and identically distributed (i.i.d.) additive errors 
\begin{align}
\label{eq:def_shift_errors}
    \hat{\boldsymbol{x}}\rightarrow\hat{\boldsymbol{x}}' \equiv \hat{\boldsymbol{x}} + \boldsymbol{\xi},
\end{align}
where $\boldsymbol{\xi}\equiv(\xi_q^{(1)}, \xi_q^{(1)}, ..., \xi_q^{(N)}, \xi_q^{(N)})\sim_\text{i.i.d.}\mathcal{N}(0,\sigma^2)$ are the random shifts that follow an i.i.d. Gaussian distribution. In other words, the probability density of $\boldsymbol{\xi}$ is given by
\begin{align}
\label{eq:def_Gaussian}
    P_\sigma(\boldsymbol{\xi}) \equiv \frac{1}{\sqrt{2\pi\sigma^2}}\exp\left(-\frac{||\boldsymbol{\xi}||^2}{2\sigma^2}\right).
\end{align}
By measuring the stabilizers, we obtain $\boldsymbol{s}\in\mathbb{R}^{2N}$, a set of real-valued syndromes.
Since the true error $\boldsymbol{\xi}$ is not known a priori, we assume $\boldsymbol{\eta}_{\boldsymbol{s}}$ to be a candidate error that is consistent with the syndrome $\boldsymbol{s}$ (for more details on the definitions of syndrome and candidate errors, we refer the readers to Sec. IV A of Ref.~\cite{lin2023closest}).
With these, MLD aims to find the logical operator $\boldsymbol{l}^\perp$ that maximizes the following coset probability \cite{conrad2022gottesman}
\begin{align}
\label{eq:coset_prob}
Z(\boldsymbol{\eta_s}, \boldsymbol{l}^\perp)\equiv\sum_{\boldsymbol{u}\in\sqrt{2\pi}\Lambda}P_\sigma(\boldsymbol{\eta_s}-\boldsymbol{l}^\perp-\boldsymbol{u}) .
\end{align}
After the optimal solution $\boldsymbol{l}_*^\perp$ is found, the corresponding displacement operator will be applied to attempt to correct the random shift error $\boldsymbol{\xi}$. Note that the MLD is an optimal strategy to minimize the chance of getting a logical error after the attempted shift correction \cite{bravyi2014efficient}. 

\section{More details on the exact MLD for the surface-square GKP codes}
\label{app: Review the BSV decoders}

In this section, we review the exact MLD for the ``unrotated'' surface code, proposed by Bravy, Suchara and Vargo (BSV) in Ref.~\cite{bravyi2014efficient}. We will explain how to adapt this method to perform an efficient and exact MLD on a ``rotated'' surface-square GKP code.

\begin{figure}[!ht]
\centering
\includegraphics[width=\linewidth]{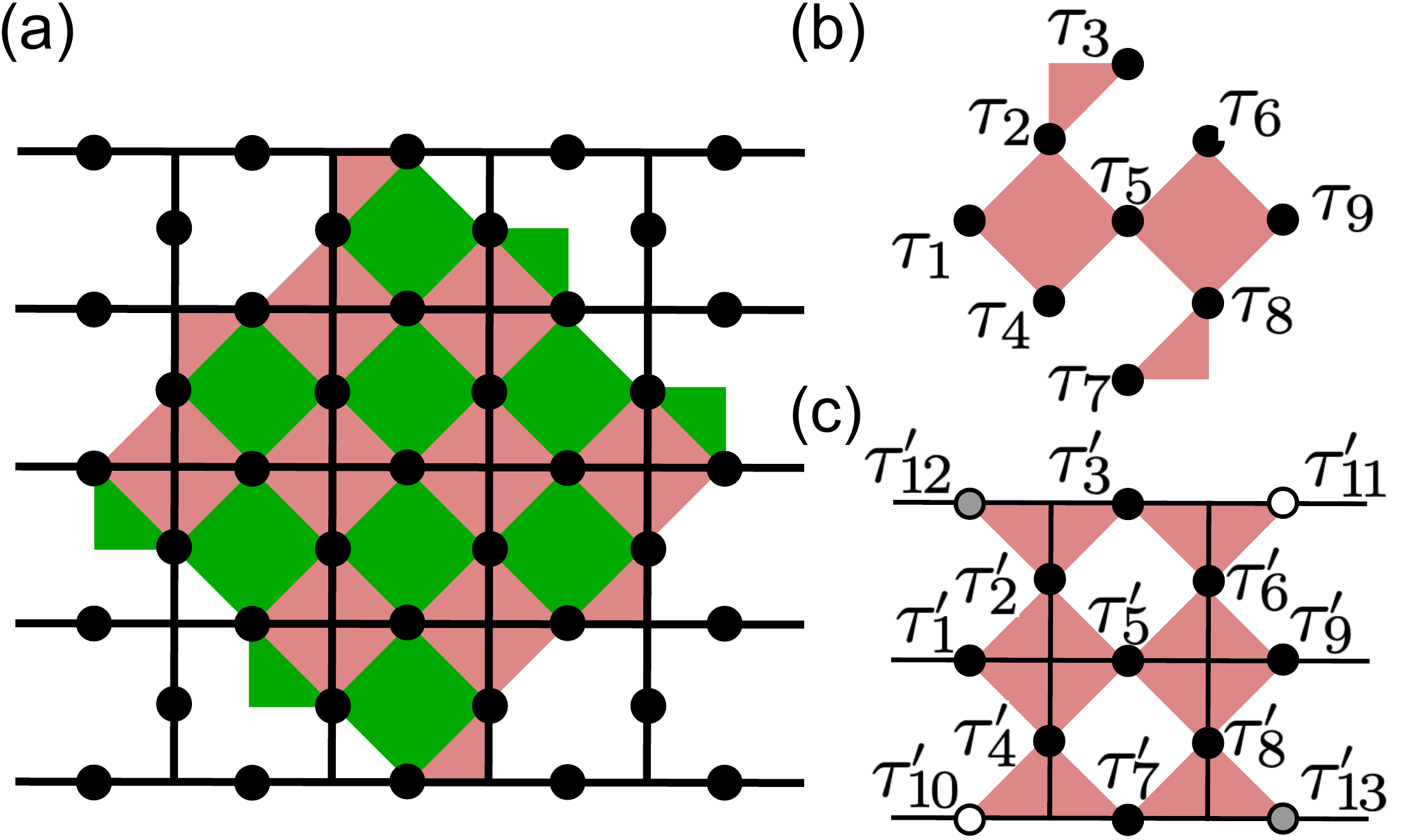}
 \caption{Embedding a ``rotated'' surface-square GKP code into an ``unrotated'' surface code.
 (a) An example embedding for the $d=5$ surface-square GKP code.
 (b) The $X$-type stabilizers for the $d=3$ surface-square GKP code. Each GKP qubit has weight $\tau_i = \tau_\text{sq}(g_i=1)/\tau_\text{sq}(g_i=0)$ where $\tau_\text{sq}(g_i)$ is defined in Eq.~\eqref{eq:def_tau}.
 (c) The  $X$-type stabilizers for the $d=3$ ``unrotated'' surface code. For the data qubits shown in black, they correspond to the GKP qubits in the surface-square GKP code with weights $\tau'_i\equiv \tau_i$. For the data qubits shown in white and grey, their weights are set to 0 and 1 respectively, such that the coset probability for the ``unrotated'' surface code correspond to that for the surface-square GKP code.  
}
 \label{fig:bsv}
\end{figure}

In contrast to the rotated surface-square GKP code considered in our work, Ref.~\cite{bravyi2014efficient} considered the surface code in a different geometry, which is often referred to as the  ``unrotated'' surface code. 
In Fig.~\ref{fig:bsv}(a), we show an example of $d=5$ ``unrotated'' surface code where the black dots represent the data qubits in the code. 
In the ``unrotated'' surface code, the $X$-type stabilizers (red) are often referred to as the site stabilizers, which act on four data qubits of the vertices, and the $Z$-type stabilizers (green) are the plaquette stabilizers, which act on four data qubits of the square plaquettes.
Stabilizers of either type that are on the boundaries act on only three data qubits. In our work, to study ``rotated'' surface-square GKP codes, we embed a ``rotated'' surface-square GKP code into an ``unrotated'' surface code as shown in Fig.~\ref{fig:bsv}(a). 
This is the starting point of applying the BSV decoder to surface-square GKP codes.

One of the seminal contributions in Ref.~\cite{bravyi2014efficient} is a method for calculating the coset probabilities exactly and efficiently for the ``unrotated'' surface codes subject to a pure $Z$-noise or $X$-noise model. More specifically, there are $N'\equiv d^2+(d-1)^2$ data qubits for the distance-$d$ ``unrotated'' surface code and we assign to each data qubit a weight $\tau'_i\in\mathbb{R}$ for $1\leq i\leq N'$. Then, let $\boldsymbol{g}'$ denote a binary vector with $N'$ components describing an $X$-type stabilizer of the ``unrotated'' surface code. 
The coset probabilities for the ``unrotated'' surface code can then be written in the following form for a given set of weights $\boldsymbol{\tau}'$,
\begin{align}
\label{eq:def_Z_w}
    Z(\boldsymbol{\tau}') \equiv\sum_{\boldsymbol{g}'\in\mathcal{G}_X'}\prod_{i=1}^{d^2+(d-1)^2}(\tau'_i)^{g'_i},
\end{align}
where $\mathcal{G}_X'$ is the set of all the $X$-type stabilizers of the ``unrotated'' surface code, and $g'_i$  denotes the $i$-th component of $\boldsymbol{g}'$. Note that $(\tau'_i)^{g'_i}=\tau'_i$ if $g'_i=1$ and $(\tau'_i)^{g'_i}=1$ if $g'_i=0$.
Notably, $Z(\boldsymbol{\tau}')$ can be calculated efficiently in time $O(d^4)$ (see Sec. V of Ref.~\cite{bravyi2014efficient} for more details). Upon adequate adaptations, this technique can be used to efficiently decode a surface-square GKP code as we describe below. 

Because we are interested in a noise model where the random shifts in the $\hat{q}$ and $\hat{p}$ subspaces are independent from each other, we exclusively consider the lattice in the $N$-dimensional $\hat{p}$ subspace $(N=d^2)$, denoted as  $\Lambda^{(\hat{p})}_\text{surf-sq}$. Similar analysis can be carried out for the $\hat{q}$ subspace as well. For this noise model, Eq.~\eqref{eq:coset_prob} factorizes to a product of two summations $Z(\boldsymbol{\eta_s}, \boldsymbol{l}^\perp)=Z(\boldsymbol{\eta_s}^{(\hat{q})}, \boldsymbol{l}^{\perp,(\hat{q})})Z(\boldsymbol{\eta_s}^{(\hat{p})}, \boldsymbol{l}^{\perp,(\hat{p})})$ where
\begin{align}
\label{eq:coset_prob_p}
Z(\boldsymbol{\eta_s}^{(\hat{p})}, \boldsymbol{l}^{\perp,(\hat{p})})=\sum_{\boldsymbol{u}\in\sqrt{2\pi}\Lambda^{(\hat{p})}_\text{surf-sq}}P_\sigma(\boldsymbol{\eta_s}^{(\hat{p})}-\boldsymbol{l}^{\perp,(\hat{p})}-\boldsymbol{u}) .
\end{align}
Similar expression holds for $Z(\boldsymbol{\eta_s}^{(\hat{q})}, \boldsymbol{l}^{\perp,(\hat{q})})$.

As shown in Ref.~\cite{lin2023closest} (see Sec. IX therein), $\sqrt{2\pi}\Lambda^{(\hat{p})}_\text{surf-sq}$ contains $2\sqrt{\pi}Z_{N}$ as a sublattce, which allows us to classify the lattice points in  $\sqrt{2\pi}\Lambda^{(\hat{p})}_\text{surf-sq}$ to different equivalent classes. In particular, for any $\boldsymbol{u}\in\sqrt{2\pi}\Lambda^{(\hat{p})}_\text{surf-sq}$, it can be written as $\boldsymbol{u}=\sqrt{\pi}(\boldsymbol{g}+\boldsymbol{v})$ where $\boldsymbol{g}$ is the binary vector for a $X$-type stabilizer, and $\boldsymbol{v}\in2Z_{N}$. Then defining 
\begin{align}
    \label{eq:def_x}
    \boldsymbol{\eta}' &\equiv\frac{1}{\sqrt{\pi}}(\boldsymbol{\eta_s}^{(\hat{q})}-\boldsymbol{l}^{\perp,(\hat{q})}),
\end{align} 
then the summation in Eq.~\eqref{eq:coset_prob_p} can be written as 
\begin{equation}
\label{eq:MLD_5}
\begin{aligned}
    &\sum_{\boldsymbol{u}\in\sqrt{2\pi}\Lambda^{(\hat{p})}_\text{surf-sq}}P_\sigma(\sqrt{\pi}\boldsymbol{\eta}'-\boldsymbol{u})\\
    =&\sum_{\boldsymbol{g}\in\mathcal{G}_X}\sum_{\boldsymbol{v}\in2Z_N}P_\sigma(\sqrt{\pi}(\boldsymbol{\eta}'-\boldsymbol{g}-\boldsymbol{v}))\\
    =&C\sum_{\boldsymbol{g}\in\mathcal{G}_X}\sum_{\boldsymbol{v}\in2Z_N}\prod_{i=1}^{N}\exp\left(-\frac{\pi(\eta'_{i}-g_i-v_i)^2}{2\sigma^2}\right)\\
    =&C\sum_{\boldsymbol{g}\in\mathcal{G}_X}\prod_{i=1}^{N}\tau_\text{sq}(\eta'_{i}, g_i, \sigma),
\end{aligned}
\end{equation}
where $C=(2\pi\sigma^2)^{-N/2}$ is an unimportant prefactor from the Gaussian distribution which will be omitted hereafter, and
\begin{align}
\label{eq:def_tau}
    \tau_\text{sq}(\eta'_{i}, g_i, \sigma) \equiv \sum_{{n}\in\mathbb{Z}}\exp\left(-\frac{\pi(\eta'_{i}-g_i-2n)^2}{2\sigma^2}\right).
\end{align}
Here $\mathcal{G}_X$ denotes the group of $X$-type stabilizers. Equations similar to Eqs.~\eqref{eq:MLD_5}-\eqref{eq:def_tau} for general concatenated-square GKP codes have been obtained in Ref.~\cite{conrad2022gottesman}, but the authors did not investigate further the performance of the MLD for explicitly GKP codes. 
Below we will explain how to use the technique in Ref.~\cite{bravyi2014efficient} to calculate the quantity in Eqs.~\eqref{eq:MLD_5}-\eqref{eq:def_tau} for the specific case of the surface-square code, given its similarity to $Z(\boldsymbol{\tau}')$ in Eq.~\eqref{eq:def_Z_w}. 

Despite the similarities, there are two critical differences between Eq.~\eqref{eq:def_Z_w} and Eq.~\eqref{eq:MLD_5}. First, each summand in $Z(\boldsymbol{\tau}')$ has the same number of factors of $\tau_i'$ as the hamming weight of the corresponding stabilizer, whereas the summands in Eq.~\eqref{eq:MLD_5} always have $N$ factors of $\tau_\text{sq}(\eta'_{i}, g_i, \sigma)$.
This is due to the fact that $\tau_\text{sq}(\eta'_{i}, g_i=0, \sigma)\neq 1$, as defined in Eq.~\eqref{eq:def_tau}. To resolve this difference, we notice
\begin{align}
\label{eq:def_w_i_g_i}
    \tau_\text{sq}(g_i) = \left(\frac{\tau_\text{sq}(g_i=1)}{\tau_\text{sq}(g_i=0)}\right)^{g_i}\tau_\text{sq}(g_i=0),
\end{align}
for both $g_i\in\lbrace0,1\rbrace$.
Hereafter, we will use $\tau_\text{sq}(g_i)\equiv\tau_\text{sq}(\eta'_{i}, g_i; \sigma)$ for simplicity.
With that, the quantity in Eq.~\eqref{eq:MLD_5} can be rewritten as
\begin{equation}
\label{eq:MLD_7}
\begin{aligned}
    \prod_{i=1}^{N}\left(\tau_\text{sq}(g_i=0)\right)\sum_{\boldsymbol{g}\in\mathcal{G}_X}\prod_{i=1}^{N}\left(\frac{\tau_\text{sq}(g_i=1)}{\tau_\text{sq}(g_i=0)}\right)^{g_i},
\end{aligned}
\end{equation}
where the summand now has exactly the same form as that in $Z(\boldsymbol{\tau}')$.
Since the first product in Eq.~\eqref{eq:MLD_7} is easy to calculate, we will focus on the summation  in the quantity.
Second, we further note that $Z(\boldsymbol{\tau}')$ has more summands than that in Eq.~\eqref{eq:MLD_7}, due to the fact that ``unrotated'' surface code has more data qubits than the surface-square GKP code. 
To resolve this difference, we have to set certain weights in the ``unrotated'' surface code to be either 0 or 1. 
As an example, in Fig.~\ref{fig:bsv}(b) we show the $X$-type stabilizers for the $d=3$ surface-square GKP code together with its weights 
$$
\tau_i\equiv \tau_\text{sq}(g_i=1)/\tau_\text{sq}(g_i=0).
$$
In order to calculate the quantity in Eq.~\eqref{eq:MLD_7}, we embed the surface-square GKP code into the $d=3$ ``unrotated'' surface code as shown in Fig.~\ref{fig:bsv}(c), where $\tau'_i=\tau_i$ for $1\leq i\leq d^2$. In such embedding, we notice that the hamming weights of boundary stabilizers in the surface-square GKP codes have increased from 2 to 3. 
For codes with larger distances, the hamming weights of boundary stabilizers may be increased to 4, see for example Fig.~\ref{fig:bsv}(a).
For that, we shall set $\tau'_{13}=\tau'_{12}=1$ (colored in grey) such that the presence of these two additional data qubits do not affect the value of the corresponding product.
Similarly, we notice there are two additional unwanted boundary stabilizers in the ``unrotated'' surface code at the top-right and left-bottom corners. For that, we shall set $\tau'_{10}=\tau'_{11}=0$ (colored in white) to suppress their contribution to the quantity of interest.
More generally, in order to evaluate the quantity in Eq.~\eqref{eq:MLD_7} for the $X$-type stabilizers, for the $(2d-4)$ qubits in the top-left and bottom-right corners that are adjacent to the embedded surface-square GKP code, we will set their weights to one, and for the other $\frac{(d-1)^2}{2}-(2d-4)$ qubits in the top-left and bottom-right corners, together with all the $\frac{(d-1)^2}{2}$ qubits in the top-right and bottom-left corners, their weights will be set to zero.
For evaluating the quantity in Eq.~\eqref{eq:MLD_7} for the $Z$-type stabilizers, the easiest approach is to rotate the stabilizers and qubits in the surface-square GKP code by 90 degrees, then we can follow the same approach as outlined above.

Let us make a remark for the weights $\tau_\text{sq}(g_i)$ defined in Eq.~\eqref{eq:def_w_i_g_i}. Since they are the inputs for the BSV decoder, their accuracy is critical for the performance of the BSV decoder. Because
\begin{align}
    \tau_\text{sq}(g_i) = \sum_{{n}\in\mathbb{Z}}\exp\left(-\frac{2\pi}{\sigma^2}\left(\frac{\eta'_i-g_i}{2}-n\right)^2\right),
\end{align}
it suggests that we can simply find the closest integers for $(\eta'_i-g_i)/2$ in order to estimate $\tau_\text{sq}(g_i)$. 
We notice that for $\sigma=0.6$,
\begin{align}
    \exp\left(-\frac{2\pi}{\sigma^2}\tilde{n}^2\right)
\end{align}
is of the order $10^{-31}$ and $10^{-69}$ for $\tilde{n}=2$ and $3$ respectively. Hence, for the $i$-th GKP qubit, we only need to find $N_v$ closest integers, where $N_v\lesssim 4$, to accurately approximate its weight, which results in $O(N_vd^2)$ runtime complexity to prepare the inputs for the BSV decoder. 

\section{More details on the tensor-network decoder for the color-GKP codes}
\label{sec: More details on the tensor-network decoder for the color-GKP codes}

In this section, we provide more details for the tensor-network MLD for the color-GKP codes. 
Similar to Eq.~\eqref{eq:MLD_5} for the surface-square code, the MLD for the color-square code aims to evaluate
\begin{equation}
\label{eq:MLD_8}
\begin{aligned}
    \sum_{\boldsymbol{u}\in\sqrt{2\pi}\Lambda^{(\hat{p})}_\text{color-sq}}P_\sigma(\boldsymbol{\eta_s}-\boldsymbol{l}^\perp-\boldsymbol{u}) = \sum_{\boldsymbol{g}\in\mathcal{G}_X}\prod_{i=1}^{N}\tau_\text{sq}(\eta'_{i}, g_i, \sigma),
\end{aligned}
\end{equation}
where $\tau_\text{sq}(\eta'_{i}, g_i, \sigma)$ is defined in Eq.~\eqref{eq:def_tau}, and $\mathcal{G}_X$ is the $X$-type stabilizers for the color-square GKP code.
Here we consider the $N$-dimensional $\hat{p}$ subspace of the color-square code because the random shifts in the $\hat{q}$ and $\hat{p}$ subspaces are independent from each other. 
This is, however, not the case for the color-hexagonal code, and we need to consider the full $2N$-dimensional lattice $\Lambda_\text{color-hex}$. As one can show, $\sqrt{2\pi}\Lambda_\text{color-hex}$ contains a sublattice generated by the transpose of $\tilde{S}\equiv\sqrt{\pi}\oplus_{i=1}^NS$, where $S$ is the symplectic matrix for the one-mode hexagonal code. 
Hence for any $\boldsymbol{u}\in\sqrt{2\pi}\Lambda_\text{color-hex}$, it can be written as $\boldsymbol{u}=\tilde{S}(\boldsymbol{g}+\boldsymbol{v})$ where $\boldsymbol{g}$ is the binary vector for certain stabilizer, and $\boldsymbol{v}\in2Z_{2N}$.
Similar to Eq.~\eqref{eq:def_x}, we define 
\begin{align}
    \label{eq:eff_can_error}
    \boldsymbol{\eta_s}-\boldsymbol{l}^\perp=\tilde{S}\boldsymbol{\eta},
\end{align}
then the summation in Eq.~\eqref{eq:coset_prob} reduces to 
\begin{equation}
\label{eq:MLD_5_app}
    \begin{aligned}
    &\sum_{\boldsymbol{g}\in\mathcal{G}}\sum_{\boldsymbol{v}\in2Z_{2N}}P_\sigma(\tilde{S}(\boldsymbol{\eta}-\boldsymbol{g}-\boldsymbol{v}))\\
    =&\sum_{\boldsymbol{g}\in\mathcal{G}}\sum_{\boldsymbol{v}\in2Z_{2N}}\prod_{i=1}^{N}P_\sigma(\sqrt{\pi}S[\boldsymbol{\eta}-\boldsymbol{g}-\boldsymbol{v}]_i)\\
    =&\sum_{\boldsymbol{g}\in\mathcal{G}}\prod_{i=1}^{N}\tau(S, [\boldsymbol{\eta}]_i, [\boldsymbol{g}]_i , \sigma),
    \end{aligned}
\end{equation}
where $\mathcal{G}$ is the full stabilizer group, and 
\begin{equation}
\label{eq:def_tau_conc}
    \begin{aligned}
    \tau(S, [\boldsymbol{\eta}]_i, [\boldsymbol{g}]_i , \sigma) &\equiv \sum_{\boldsymbol{v}\in Z_2}P_\sigma(\sqrt{\pi}S([\boldsymbol{\eta}]_i-[\boldsymbol{g}]_i-2\boldsymbol{v})) .
    \end{aligned}
\end{equation}
Here we have used $[\boldsymbol{b}]_i$ to denote the two-dimensional projection of $\boldsymbol{b}$ onto the $i$-th qubit, and the summation in Eq.~\eqref{eq:def_tau_conc} is carried over the two dimensional integer lattice. 

\begin{figure}[!ht]
\centering
\includegraphics[width=\linewidth]{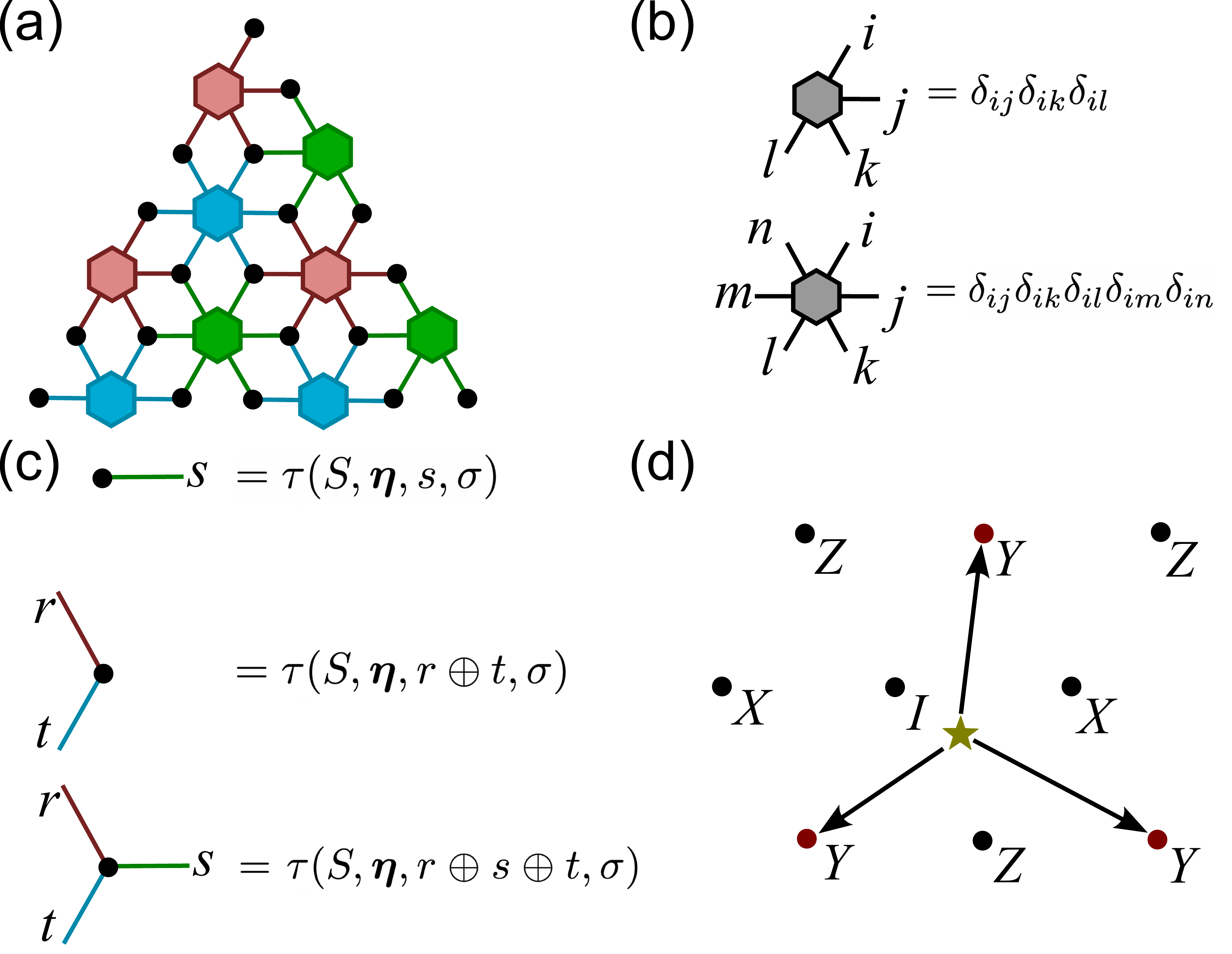}
 \caption{The tensor network MLD for the color-square and color-hexagonal GKP codes. (a) An illustration of the tensor network for $d=5$ color code. (b) Definition of all the tensors correspond to stabilizers, up to certain rotation. (c) The tensors correspond to the data qubits are defined in Eq.~\eqref{eq:def_tau} and Eq.~\eqref{eq:def_tau_conc} for the color-square and color-hexagonal codes respectively. See the description in the text.
 For both types of tensors, corresponding to either stabilizers or data qubits, their bond dimension is 2 and 4 for the color-square and color-hexagonal codes respectively. (d) Illustration of the lattice $\Lambda(2\sqrt{\pi}S^T)$ where $S$ is the symplectic matrix corresponds to the one-mode hexagonal code. Each lattice point corresponds to a logical error for the qubit. The probability of a logical error, say $Y$, is a summation of the contributions from the corresponding lattice points, which depends on their distance to the syndrome (brown star) and the noise strength $\sigma$. Note that we have only illustrated the contributions from three lattice points, and the contributions from other lattice points can be neglected because the probability decays exponentially with respect to the distance squared. 
}
 \label{fig:tensor_network_color_GKP_codes}
\end{figure}

We remark that Eqs.~\eqref{eq:MLD_5_app}-\eqref{eq:def_tau_conc} is applicable for generic concatenated-GKP code.
Because $S=I_2$ for the concatenated-square GKP code, Eq.~\eqref{eq:def_tau_conc} reduces to a product of two summations for the $\hat{q}$ and $\hat{p}$ subspaces, where each summation has the same form as Eq.~\eqref{eq:MLD_8}.
In other word, Eqs.~\eqref{eq:MLD_5_app}-\eqref{eq:def_tau_conc} represent a straightforward generalization for the result obtained in Ref.~\cite{conrad2022gottesman}, for approximating MLD for generic concatenated-GKP codes.

The tensor-network method to calculate the quantities in Eq.~\eqref{eq:MLD_8} and Eq.~\eqref{eq:MLD_5_app} was first described in Ref.~\cite{bravyi2014efficient} for decoding surface code, and later generalized to general 2D stabilizer codes in Ref.~\cite{chubb2021general}.
The tensor network for the color-square code is depicted in Fig.~\ref{fig:tensor_network_color_GKP_codes}(a) which 
typically involves two types of tensors. The first type of tensors, shown in Fig.~\ref{fig:tensor_network_color_GKP_codes}(b), correspond to stabilizers and take value one iff all the indices of the tensor have the same value, and zero otherwise.
The second type of tensors, shown in Fig.~\ref{fig:tensor_network_color_GKP_codes}(c), correspond to qubits which, as defined in Eq.~\eqref{eq:def_tau}, depend on the values of the candidate error $\boldsymbol{\eta}_{\boldsymbol{s}}$, the logical error $\boldsymbol{l}^\perp$ and the stabilizer $\boldsymbol{g}$ at the qubit. 
Taking the rank-3 tensor in Fig.~\ref{fig:tensor_network_color_GKP_codes}(c) as an example, we have $g_i=r\oplus s\oplus t$ where $r,s,t$ are the indices of the tensor and $g_i,r,s,t\in\mathbb{F}_2\equiv\left\{0,1\right\}$.
For the rest of this section, we use $\oplus$ to denote the addition in $\mathbb{F}_2$ or $\mathbb{F}_4$, see below.
We note that $g_i$ is essentially the local logical error for the $i$-th qubit, if we view $\boldsymbol{\eta}'$, defined in Eq.~\eqref{eq:def_x}, as an effective candidate error.

For color-hexagonal code, the tensor network has the same topology as the one for the color-square code, but the tensors are defined differently. In particular, because both $X$ and $Z$ type stabilizers are needed to decode the color-hexagonal code, and the two types of stabilizers can be viewed lying on top of each other geometrically,
the tensors in Fig.~\ref{fig:tensor_network_color_GKP_codes}(b) correspond to the composition of both types of stabilizers.
Taking the rank-4 tensor in Fig.~\ref{fig:tensor_network_color_GKP_codes}(b) as an example, its indices take values in $\mathbb{F}_4\equiv\left\{0,1,2,3\right\}$ which corresponds to $I$, $X$, $Z$ and $Y$ for the composite stabilizer.
Similarly, the tensors for the qubits have bond dimension 4 with values given in Eq.~\eqref{eq:def_tau_conc}.
In particular, because $[\boldsymbol{g}]_i=\left\{0,1\right\}^2$, it can be mapped to an $g_i\in\mathbb{F}_4$ which is essentially the local logical error for the $i$-th qubit, if we view $\boldsymbol{\eta}$, defined in Eq.~\eqref{eq:eff_can_error}, as the effective candidate error. 
Take the rank-3 tensor in Fig.~\ref{fig:tensor_network_color_GKP_codes}(c) as an example, we have $g_i=r\oplus s\oplus t$ where $r,s,t$ are the indices of the tensor and $g_i,r,s,t\in\mathbb{F}_4$.
The tensors shown in Fig.~\ref{fig:tensor_network_color_GKP_codes}(c) are defined in Eq.~\eqref{eq:def_tau_conc} with square brackets and subscripts omitted for simplicity.
The method of calculating Eq.~\eqref{eq:def_tau_conc} will be described below in the context of brute force MLD for the $[[5,1,3]]$-hexagonal GKP code.

\section{More details on the brute force MLD for the $[[5,1,3]]$-hexagonal GKP code}
\label{sec:More details on the brute force MLD for the 513-hexagonal GKP code}

\begin{table}[h!]
\centering
\begin{tabular}{| c |} 
 \hline
  \mybox[0.25cm]{$I$}\mybox[0.25cm]{$X$}\mybox[0.25cm]{$Z$}\mybox[0.25cm]{$Z$}\mybox[0.25cm]{$X$} \\
  \mybox[0.25cm]{$X$}\mybox[0.25cm]{$I$}\mybox[0.25cm]{$X$}\mybox[0.25cm]{$Z$}\mybox[0.25cm]{$Z$} \\
  \mybox[0.25cm]{$Z$}\mybox[0.25cm]{$X$}\mybox[0.25cm]{$I$}\mybox[0.25cm]{$X$}\mybox[0.25cm]{$Z$} \\
  \mybox[0.25cm]{$Z$}\mybox[0.25cm]{$Z$}\mybox[0.25cm]{$X$}\mybox[0.25cm]{$I$}\mybox[0.25cm]{$X$} \\
\hline       
\end{tabular}
\caption{The stabilizer generators for the [[5,1,3]] code. 
}
\label{tab:stabilizers_513}
\end{table}

In this section, we provide more details on the brute-force MLD for the $[[5,1,3]]$-hexagonal GKP code. The brute-force MLD aims to evaluate exactly the same expressions in Eq.~\eqref{eq:MLD_5_app}-\eqref{eq:def_tau_conc}, except that $\mathcal{G}\equiv\mathcal{G}_{513}$ is for the full stabilizer group of the $[[5,1,3]]$ code. We have listed the generators for $\mathcal{G}_{513}$ in Tab.~\ref{tab:stabilizers_513}. In total there are $16$ elements in $\mathcal{G}_{513}$ which makes the brute-force MLD feasible. 
As an example, let us consider the summand for the stabilizer $IYXXY$ which is a product of five terms, each has the same form as that in Eq.~\eqref{eq:def_tau_conc}.
Taking second qubit as an example, because the second Pauli operator of the stabilizer is $Y$, the corresponding $[\boldsymbol{g}]_{i=2}=[1,1]^T$ using our convention.
With that, we can plug each $[\boldsymbol{g}]_i$ into Eq.~\eqref{eq:def_tau_conc} to estimate the weight in the $i$-th qubit.
Specifically, let
\begin{align}
    \boldsymbol{y} \equiv \sqrt{\pi}S([\boldsymbol{\eta}]_i-[\boldsymbol{g}]_i),
\end{align}
then the summation in Eq.~\eqref{eq:def_tau_conc} can be written as
\begin{align}
    \sum_{\boldsymbol{v}\in Z_2}P_\sigma(\boldsymbol{y}-2\sqrt{\pi}S\boldsymbol{v}) .
\end{align}
Because $2\sqrt{\pi}S\boldsymbol{v}$ can be viewed as a point in a two dimensional lattice generated by $2\sqrt{\pi}S^T$, in order to approximate the summation, we can simply find $N_v\lesssim 4$ lattice points that are closest to $\boldsymbol{y}$, and sum up their contributions to the summation. The truncation can be similarly justified as we did above for the exact MLD for surface-square GKP code.
This procedure of calculating $\tau(S, [\boldsymbol{\eta}]_i, [\boldsymbol{g}]_i , \sigma)$ is illustrated in Fig.~\ref{fig:tensor_network_color_GKP_codes}(d).

\section{More discussions on prior works}
\label{sec:Comparison to prior works}

In this section, we compare our results to several prior works \cite{fukui2017analog, fukui2018high, hanggli2020enhanced, zhang2022concatenation}, which also study the thresholds of multi-mode GKP codes near $\sigma=1/\sqrt{e}\simeq 0.6065$.

In Ref.~\cite{fukui2017analog}, the authors considered the $C_4/C_6$ code concatenated with square-GKP codes. They showed that the code has threshold between $0.60$ and $0.61$ when the analog information from the inner square GKP codes is incorporated into the decoder. The numerical evidence is shown in Fig. 3 therein, where the noise strength $\sigma$ is scanned with the resolution of $0.01$. Because of the limited resolution, it is difficult to conclude that the code has a threshold above $\sigma=1/\sqrt{e}\simeq 0.6065$.

In Ref.~\cite{fukui2018high}, the authors applied the decoding strategy in Ref.~\cite{fukui2017analog} to surface-square GKP code, and the results are presented in Fig. 2 therein. Close inspection of Fig. 2(b) reveals that surface-square GKP code at $d=25$ has higher failure probability than $d=23$ at $\sigma=0.605$ using the suboptimal decoding method therein. 
In order to resolve the ambiguities in Ref.~\cite{fukui2017analog, fukui2018high}, we have chosen to scan the noise strength with higher resolution of 0.001 in our simulations. Moreover, we have carefully studied the fidelity of the surface-square GKP code near $\sigma=1/\sqrt{e}\approx0.6065$ as shown in Fig.~\ref{fig:surface_square_performance}(b) in the main text. 
The result in Ref.~\cite{fukui2018high} is consistent with our conclusion that surface-square code can have threshold near $\sigma=1/\sqrt{e}\approx0.6065$ but only by using the optimal decoding method.

In Ref.~\cite{hanggli2020enhanced}, the authors studied the QEC capacity of several multi-mode GKP codes, including surface-rectangular and surface-asymmetric-hexagonal GKP codes. With a tensor-network decoder, they also showed that the analog information from the inner GKP codes can significantly improve the QEC capacity of these codes, and the main numerical results are shown in Fig. 8 therein. But, similar to Ref.~\cite{fukui2017analog}, the noise strength $\sigma$ is scanned only with the resolution of $0.01$, and one can only conclude that these codes have threshold between $0.60$ and $0.61$.
We also note that the decoders in Ref.~\cite{hanggli2020enhanced} are approximate MLD since a tensor-network decoder with a finite maximum bond dimension cutoff was used. However as we showed above, MLD of a surface-rectangular GKP code can in fact be performed exactly and efficiently with our adaptation of the BSV method because the shift errors in the position and momentum subspaces are decoupled from each other. Hence, the threshold of the surface-rectangular GKP code shown in Fig. 8(b) in Ref.~\cite{hanggli2020enhanced} is not necessarily optimal. It would be interesting to use our adaptation of the exact BSV method to explore the thresholds of the the surface-rectangular GKP codes with various aspect ratios, or surface-GKP codes on rectangular geometries (instead of square geometry shown in Fig.~\ref{fig:bsv}). 

In Ref.~\cite{zhang2022concatenation}, the authors studied the XZZX-rectangular GKP codes decoded by using a minimum-weight-perfect-matching decoder. In Fig. 4 therein, the authors claimed that the code has threshold of $0.67$ when the square root of the aspect ratio of the inner rectangular GKP codes is equal to $2.1$. We note, however, that the simulation was only carried out for codes up to distance $d=9$. Further, close inspection of Fig. 4(b) reveals that the crossings of the logical error rates for distances $d$ and $d+2$ \emph{decreases} as the distance is increased from $d=3$ to $d=7$. It is reasonable to suspect that the crossings could further decrease as the distance is increased further, and the threshold, a quantity defined only in the asymptotic limit of $d\rightarrow\infty$, could be much smaller than $0.67$ if the threshold ever exists for these codes. Hence, based only on the data presented, it is difficult to confidently conclude that the threshold of the XZZX-rectangular GKP code is larger than $\sigma=1/\sqrt{e} \simeq 0.6065$.

In our work, we have carefully determined the crossings of the fidelity between $d$ and $d+2$ for surface-square and color-hexagonal GKP codes, as shown in the insets of Fig.~\ref{fig:surface_square_performance}(a) and Fig.~\ref{fig:color_hex_vs_surf_sq} respectively. 
In particular, the crossings of the surface-square code gradually increase as the code distance is increased, showing that the optimal threshold of surface-square code is indeed close to $\sigma=1/\sqrt{e}\approx0.6065$. This is further confirmed by a more in-depth study of the performance at $\sigma = 1/\sqrt{e}$ as a function of code distance $d$ as shown in Fig.~\ref{fig:surface_square_performance}(b).

\section{More details on the hashing bound of Pauli channels}
\label{sec: More details on the hashing bound of Pauli channels}

In this section, we provide more details on how to use the hashing bound of Pauli channels to compute the achievable rate of the GKP codes considered. In this work, we consider encoding a single qubit using $N$-mode GKP codes, which simulates the single-qubit Pauli error channel $\mathcal{P}_{\vec{p}}$ with $N$ independent copies of Gaussian random displacement channel. As a result, the achievable rate of the GKP code is that of the Pauli channel divided by $N$, i.e.,
\begin{align}
\label{eq:achievable_rate_GKP_as_Pauli_divided_N}
    R = \frac{1}{N}\text{max}_{\hat{\rho}}(I_c(\mathcal{P}_{\vec{p}}, \hat{\rho}))
\end{align}
where 
\begin{align}
\label{eq:def_Ic_pauli}
I_{c}(\mathcal{P}_{\vec{p}},\hat{\rho}) \equiv S(\mathcal{P}_{\vec{p}}(\hat{\rho})) - S(\mathcal{P}_{\vec{p}}^{c}(\hat{\rho}))
\end{align}
is the coherent information of the Pauli channel $\mathcal{P}_{\vec{p}}$ with respect to an input state $\hat{\rho}$. By definition, we have
\begin{align}
\label{eq:def_pauli_channel}
    \mathcal{P}_{\vec{p}}(\hat{\rho}) \equiv p_I\hat{\rho} + p_X\hat{X}\hat{\rho}\hat{X} + p_Y\hat{Y}\hat{\rho}\hat{Y} + p_Z\hat{Z}\hat{\rho}\hat{Z}
\end{align}
where $\hat{X}, \hat{Y}, \hat{Z}$ are Pauli operators and $\vec{p}=(p_I,p_X,p_Y,p_Z)$ are parameters of the Pauli channel. To define the complementary channel for the Pauli channel, we consider the following isometric extension
\begin{align}
\label{eq:isometric_extension_pauli_channel}
    \tilde{\mathcal{P}}_{\vec{p}}(\hat{\rho}) = \sum_{\mu,\nu=I,X,Y,Z}\sqrt{p_\mu p_\nu}(\hat{\mu}\hat{\rho}\hat{\nu})\otimes(|\mu\rangle\langle \nu|),
\end{align}
where we identify $|\mu\rangle$ and $|\nu\rangle$ with the following basis states
\begin{equation*}
    \begin{aligned}
        |I\rangle &= (1,0,0,0)^T,\quad
        |X\rangle = (0,1,0,0)^T,\\
        |Y\rangle &= (0,0,1,0)^T,\quad
        |Z\rangle = (0,0,0,1)^T.
    \end{aligned}
\end{equation*}
We should treat $|\mu\rangle\langle \nu|$ in Eq.~\eqref{eq:isometric_extension_pauli_channel} as an environmental degree of freedom, and upon tracing it out, we recover the Pauli channel. The complementary channel is obtained by  tracing out the first degree of freedom of $\tilde{\mathcal{P}}_{\vec{p}}(\hat{\rho})$, which reads \cite{leung2017on}
\begin{widetext}
\begin{align}
\label{eq:def_pauli_channel_complementary}
    \mathcal{P}^c_{\vec{p}}(\hat{\rho}) = \begin{bmatrix}
        p_I & \sqrt{p_Ip_X}\text{Tr}(\hat{\rho}\hat{X}) & \sqrt{p_Ip_Y}\text{Tr}(\hat{\rho}\hat{Y})& \sqrt{p_Ip_Z}\text{Tr}(\hat{\rho}\hat{Z})\\
        \sqrt{p_Ip_X}\text{Tr}(\hat{\rho}\hat{X}) & p_X & -i\sqrt{p_Xp_Y}\text{Tr}(\hat{\rho}\hat{Z}) & i\sqrt{p_Xp_Z}\text{Tr}(\hat{\rho}\hat{Y})\\
        \sqrt{p_Ip_Y}\text{Tr}(\hat{\rho}\hat{Y})& i\sqrt{p_Xp_Y}\text{Tr}(\hat{\rho}\hat{Z})& p_Y & -i\sqrt{p_Yp_Z}\text{Tr}(\hat{\rho}\hat{X})\\
        \sqrt{p_Ip_Z}\text{Tr}(\hat{\rho}\hat{Z}) & -i\sqrt{p_Xp_Z}\text{Tr}(\hat{\rho}\hat{Y})& i\sqrt{p_Yp_Z}\text{Tr}(\hat{\rho}\hat{X}) & p_Z\\
    \end{bmatrix}.
\end{align}
\end{widetext}
With the forms of $\mathcal{P}_{\vec{p}}(\hat{\rho})$ and $\mathcal{P}^c_{\vec{p}}(\hat{\rho})$ obtained, we can explicitly calculate $\tilde{\mathcal{P}}_{\vec{p}}(\hat{\rho})$ for an input $\hat{\rho}$. Since the achievable rate in Eq.~\eqref{eq:achievable_rate_GKP_as_Pauli_divided_N} is maximizing over all the input states, we can obtain its lower bound by picking a particular state, say the maximally mixed state. By plugging $\hat{\rho}=\hat{I}/2$ into Eq.~\eqref{eq:def_Ic_pauli}, and \eqref{eq:def_pauli_channel}-\eqref{eq:def_pauli_channel_complementary}, we arrive that
\begin{equation}
\label{eq:achievable_rate_GKP_as_Pauli_divided_N_v2}
    \begin{aligned}
        R &=\frac{1}{N}\left(1 + \sum_{\mu=I,X,Y,Z}p_\mu\log_2 p_\mu\right).
    \end{aligned}
\end{equation}
The quantity in the parenthesis of Eq.~\eqref{eq:achievable_rate_GKP_as_Pauli_divided_N_v2} has been shown to be achievable, say using random stabilizer code \cite{lloyd1997capacity, Devetak2005, wilde2013quantum}, based on which we calculate the achievable rate for the GKP code.

\section{Derivation of the lower bound of the quantum capacity for a Gaussian random displacement channel}
\label{sec:derivation_lower_bound}

In this section, we derive the lower bound of $Q(\mathcal{N}_\sigma)$ as shown in Eq.~\eqref{eq:Gaussian_random_displacement_channel_capacity_bounds}. Recall that (see Eq.~\eqref{eq:regularized_coherent_information})
\begin{align}
    Q(\mathcal{N}_\sigma) = \lim_{N\rightarrow \infty} \frac{1}{N} \max_{\hat{\rho}} I_{c}(\mathcal{N}_\sigma^{\otimes N}, \hat{\rho}),
\end{align}
where
\begin{align}
\label{eq:def_Ic}
I_{c}(\mathcal{N}_\sigma,\hat{\rho}) \equiv S(\mathcal{N}_\sigma(\hat{\rho})) - S(\mathcal{N}_\sigma^{c}(\hat{\rho}))
\end{align}
is the coherent information of the Gaussian random displacement channel $\mathcal{N}_\sigma$ with respect to an input state $\hat{\rho}$. Here
$S(\hat{\rho}) \equiv -\text{Tr}[\hat{\rho}\log_{2}\hat{\rho}]$ is the quantum von Neumann entropy and $\mathcal{N}_\sigma^{c}$ is the complementary channel of $\mathcal{N}_\sigma$. Because coherent information might be superadditive, i.e., $Q(\mathcal{N}_1\otimes\mathcal{N}_2)\geq Q(\mathcal{N}_1) + Q(\mathcal{N}_2)$, we can lower bound $Q(\mathcal{N}_\sigma)$ using the one-shot coherent information as
\begin{align}
\label{eq:lower_bound_1}
    \max_{\hat{\rho}} I_{c}(\mathcal{N}_\sigma, \hat{\rho}) \leq Q(\mathcal{N}_\sigma).
\end{align}
Further, since the left hand side of Eq.~\eqref{eq:lower_bound_1} is a maximization over all input states, by picking a particular input state, say the single-mode thermal state $\hat{\rho}_{\bar{n}}$, we naturally obtain a lower bound for $Q(\mathcal{N}_\sigma)$. In particular, we will show that $I_{c}(\mathcal{N}_\sigma, \hat{\rho}_{\bar{n}})$ is monotonically increasing with ${\bar{n}}$ such that
\begin{align}
\label{eq:limit_Ic_thermal_state}
    \lim_{\bar{n}\rightarrow\infty} I_{c}(\mathcal{N}_\sigma, \hat{\rho}_{\bar{n}}) = \log_2\left(\frac{1}{e\sigma^2}\right).
\end{align}
This is the lower bound shown in Eq.~\eqref{eq:Gaussian_random_displacement_channel_capacity_bounds} in the main text, since $Q(\mathcal{N}_\sigma)$ is always positive.
We now describe how to prove Eq.~\eqref{eq:limit_Ic_thermal_state}.

\subsection{Preliminary}

In this subsection, we provide some background information which will be needed for later discussions. More details can be found in \cite{weedbrook2012gaussian} (with a different convention compared to the one used below).

A $N$-mode bosonic mode is associated with the $2N$ annihilation and creation operators $\hat{a}_i$ and $\hat{a}_i^\dagger$ $(1\leq i\leq N)$ satisfying $[\hat{a}_i, \hat{a}_j]=[\hat{a}_i^\dagger, \hat{a}_j^\dagger]=0$ and $[\hat{a}_i, \hat{a}_j^\dagger]=\delta_{ij}$. Another useful representation is the  position and momentum quadrature operators $\hat{\mathbf{x}}_i\equiv(\hat{q}_1, \hat{p}_1,\cdots\hat{q}_N, \hat{p}_N)^T$ where $\hat{q}_i\equiv\frac{1}{\sqrt{2}}(\hat{a}_i+\hat{a}_i^\dagger)$ and $\hat{p}_i\equiv\frac{i}{\sqrt{2}}(\hat{a}_i^\dagger-\hat{a}_i)$. The communication relation translates to $[\hat{x}_i,\hat{x}_j]=i\Omega_{ij}$ where 
\begin{align}
    \mathbf{\Omega} = \mathbf{I}_N\otimes\begin{bmatrix}
        0 & 1\\
        -1 & 0
    \end{bmatrix}.
\end{align}
Here $\mathbf{I}_N$ is the $N\times N$ identity matrix. We will also need the displacement operator defined as
\begin{align}
\label{eq:def_D}
    \hat{D}(\boldsymbol{\xi})\equiv\exp\left(i\hat{\mathbf{x}}^T\mathbf{\Omega}\boldsymbol{\xi}\right),
\end{align}
where $\boldsymbol{\xi}=(\xi^{(1)}_q, \xi^{(1)}_p,\cdots\xi^{(N)}_q, \xi^{(N)}_p)^T$. For a quantum state $\rho$, its Wigner characteristic function is defined as
\begin{equation}
\label{eq:def_chi}
    \begin{aligned}
        \chi(\boldsymbol{\xi}) &\equiv \langle \hat{D}(\boldsymbol{\xi})\rangle \equiv \text{Tr}[\hat{\rho}\hat{D}(\boldsymbol{\xi})].
    \end{aligned}
\end{equation}
Using the fact that $\text{Tr}(\hat{D}(\boldsymbol{\xi}-\boldsymbol{\xi}'))=(2\pi)^N\delta(\boldsymbol{\xi}-\boldsymbol{\xi}')$, conversely, we have
\begin{align}
    \label{eq:rho_chi}
    \hat{\rho} = \frac{1}{(2\pi)^N}\int d^{2N}\boldsymbol{\xi}\chi(\boldsymbol{\xi})\hat{D}(-\boldsymbol{\xi}).
\end{align}
The Wigner function $W(\mathbf{x})$ is the Fourier transform of $\chi(\boldsymbol{\xi})$, i.e.,
\begin{align}
    W(\mathbf{x}) = \frac{1}{(2\pi)^N}\int d^{2N}\boldsymbol{\xi}\chi(\boldsymbol{\xi})\exp(-i\mathbf{x}^T\mathbf{\Omega}\boldsymbol{\xi}).
\end{align}

We say a state $\hat{\rho}$ is Gaussian iff its Wigner characteristic function and Wigner function are Gaussian, i.e.,
\begin{align}
    \chi(\boldsymbol{\xi}; \mathbf{V}, \mathbf{\bar{x}}) &= \exp\left[-\frac{1}{2}\boldsymbol{\xi}^T(\mathbf{\Omega}\mathbf{V}\mathbf{\Omega}^T)\boldsymbol{\xi} - i(\mathbf{\Omega}\mathbf{\bar{x}})^T\boldsymbol{\xi}\right],\label{eq:chi_gaussian}\\
    W(\mathbf{x}; \mathbf{V}, \mathbf{\bar{x}}) &= \frac{\exp\left[-\frac{1}{2}(\mathbf{x}-\bar{\mathbf{x}})^T\mathbf{V}^{-1}(\mathbf{x}-\bar{\mathbf{x}})\right]}{(2\pi)^N\sqrt{\det\mathbf{V}}}.\label{eq:wigner_gaussian}
\end{align}
Hence a Gaussian state is fully characterized by the first two moments defined as
\begin{align}
    \mathbf{\bar{x}} \equiv \langle \hat{\mathbf{x}}\rangle , \quad V_{ij} \equiv \frac{1}{2}\langle \left\{\hat{x}_i-\bar{x}_i, \hat{x}_j-\bar{x}_j\right\}\rangle,
\end{align}
where $\left\{\hat{A},\hat{B}\right\}=\hat{A}\hat{B}+\hat{B}\hat{A}$. A Gaussian unitary operation is defined to be operators that transform a Gaussian state as
\begin{align}
    \mathbf{\bar{x}}\rightarrow \mathbf{S}\mathbf{\bar{x}}+\mathbf{d}, \quad \mathbf{V}\rightarrow \mathbf{S}\mathbf{V}\mathbf{S}^T,
\end{align}
where $\mathbf{d}=(d_1,\cdots,d_{2N})^T$ and $\mathbf{S}$ is a $2N\times2N$ symplectic matrix satisfying $\mathbf{S}\mathbf{\Omega}\mathbf{S}^T=\mathbf{\Omega}$. 

As an example, the single-mode thermal state is a Gaussian state given by
\begin{align}
    \hat{\rho}_{\bar{n}}\equiv\sum_{n=0}^\infty\frac{\bar{n}^n}{(\bar{n}+1)^{n+1}}|n\rangle\langle n|.
\end{align}
Using Eq.~\eqref{eq:def_D}-\eqref{eq:def_chi}, and the fact that
\begin{align}
    \langle n|\hat{D}(\alpha)|n\rangle = e^{-|\alpha|^2/2}\sum_{k=0}^n\frac{(-1)^k}{k!}{{n}\choose{k}}|\alpha|^{2k},
\end{align}
we can show that the thermal state is characterized by $\mathbf{\bar{x}}=\mathbf{0}$ and $\mathbf{V}=(\bar{n}+1/2)\mathbf{I}_2$.
We can calculate its quantum entropy as follows
\begin{equation}
    \begin{aligned}
        S(\hat{\rho}_{\bar{n}}) &= -\sum_{n=0}^\infty\frac{\bar{n}^n}{(\bar{n}+1)^{n+1}}\log\left(\frac{\bar{n}^n}{(\bar{n}+1)^{n+1}}\right)\\
        &= g(\bar{n}),
    \end{aligned}
\end{equation}
where 
\begin{align}
\label{eq:def_g}
    g(n)=(n+1)\log_2(n+1)-n\log_2(n).
\end{align}
More generally, for an arbitrary $N$-mode Gaussian state $\hat{\rho}$ with covariance matrix $\mathbf{V}$, its von Neumann entropy is given by
\begin{align}
\label{eq:S_rho_general}
    S(\hat{\rho}) = \sum_{k=1}^Ng\left(\nu_k-\frac{1}{2}\right)
\end{align}
where $\pm\nu_k$ are the eigenvalues of the matrix $i\mathbf{\Omega V}$. See more details in Sec. IIC1 in \cite{weedbrook2012gaussian}.

We will also need the following two-mode Gaussian unitary, define for the $j$-th and $k$-th mode as
\begin{align}
    \text{SUM}_{j\rightarrow k} = \exp\left[-i\hat{q}_j\hat{p}_k\right].
\end{align}
Its action on the quadrature coordinates read
\begin{align*}
    \begin{bmatrix}
        \hat{q}_j\\
        \hat{p}_j\\
        \hat{q}_k\\
        \hat{p}_k\\
    \end{bmatrix}\rightarrow\begin{bmatrix}
        \hat{q}_j\\
        \hat{p}_j-\hat{p}_k\\
        \hat{q}_k+\hat{q}_j\\
        \hat{p}_k\\
    \end{bmatrix}=\begin{bmatrix}
        1 & 0 & 0 & 0\\
        0 & 1 & 0 & -1\\
        1 & 0 & 1 & 0\\
        0 & 0 & 0 & 1
    \end{bmatrix}\begin{bmatrix}
        \hat{q}_j\\
        \hat{p}_j\\
        \hat{q}_k\\
        \hat{p}_k\\
    \end{bmatrix} \equiv \mathbf{S}_\text{SUM}\begin{bmatrix}
        \hat{q}_j\\
        \hat{p}_j\\
        \hat{q}_k\\
        \hat{p}_k\\
    \end{bmatrix}.
\end{align*}
As one can show $\mathbf{S}_\text{SUM}$ is indeed a symplectic matrix. We note that $\text{SUM}_{k\rightarrow j}$ corresponds to $\mathbf{S}_\text{SUM}^T$ for the same quadrature basis vector, and $\text{SUM}^\dagger_{j\rightarrow k}$ corresponds to 
\begin{align}
    \mathbf{S}_{\text{SUM}^\dagger}  = \begin{bmatrix}
        1 & 0 & 0 & 0\\
        0 & 1 & 0 & 1\\
        -1 & 0 & 1 & 0\\
        0 & 0 & 0 & 1
    \end{bmatrix}.
\end{align}

\subsection{Calculation of $S(\mathcal{N}_\sigma(\hat{\rho}_{\bar{n}}))$}

The Gaussian random displacement channel $\mathcal{N}_\sigma$ is defined as
\begin{align}
\label{eq:def_GRD}
    \mathcal{N}_\sigma(\hat{\rho}) \equiv \int d^{2N}\boldsymbol{\xi}P_\sigma(\boldsymbol{\xi})\hat{D}(\boldsymbol{\xi})\hat{\rho}\hat{D}^\dagger(\boldsymbol{\xi}),
\end{align}
where $d\boldsymbol{\xi}=d\xi_1\cdots\xi_{2N}$ and 
\begin{align}
    P_\sigma(\boldsymbol{\xi}) = \frac{1}{(2\pi\sigma^2)^{N}}\exp\left(-\frac{\boldsymbol{\xi}^T\boldsymbol{\xi}}{2\sigma}\right)
\end{align}
is the $2N$-dimensional Gaussian distribution. For a Gaussian state $\hat{\rho}$ corresponds to $\chi(\boldsymbol{\xi}; \mathbf{V}, \mathbf{\bar{x}})$, we can calculate $\mathcal{N}_\sigma(\hat{\rho})$ explicitly. 
Upon plugging Eq.~\eqref{eq:rho_chi}-\eqref{eq:chi_gaussian} into Eq.~\eqref{eq:def_GRD}, and carrying out the $\boldsymbol{\xi}$ integral, we have 
\begin{align*}
\label{eq:entropy_gaussian_random_displacement_for_thermal_state}
    \mathcal{N}_\sigma(\hat{\rho}) = \frac{1}{(2\pi)^N}\int d^{2N}\boldsymbol{u}\chi(\boldsymbol{u}; \mathbf{V}+\sigma^2\mathbf{I}_N, \mathbf{\bar{x}})\hat{D}(-\boldsymbol{u}),
\end{align*}
which suggests that the covariance matrix is shifted by $\sigma^2\mathbf{I}_N$. Hence for a single-mode thermal state, we have $\mathcal{N}_\sigma(\hat{\rho}_{\bar{n}})=\hat{\rho}_{\bar{n}+\sigma^2}$, such that
\begin{align}
    S(\mathcal{N}_\sigma(\hat{\rho}_{\bar{n}})) = g(\bar{n}+\sigma^2).
\end{align}

\subsection{Unitary dilation of the Gaussian random displacement channel}
In order to calculate $S(\mathcal{N}_\sigma^{c}(\hat{\rho}))$ for a single-mode Gaussian state $\hat{\rho}$, we will need to introduce a complementary channel, which will be done via unitary dilation. 
We consider a three mode Gaussian state defined by
\begin{align}
    \hat{\rho}\otimes|Q\rangle\langle Q|\otimes|P\rangle\langle P|,
\end{align}
where $\hat{\rho}$ is the state to be transmitted via the Gaussian random displacement channel, and 
\begin{align}
    |Q\rangle &=\int_{-\infty}^\infty d\xi_q\sqrt{P_\sigma(\xi_q)}|\hat{q}_2=\xi_q\rangle,\\
    |P\rangle &=\int_{-\infty}^\infty d\xi_p\sqrt{P_\sigma(\xi_p)}|\hat{p}_3=\xi_p\rangle
\end{align}
are two pure states. Here $P_\sigma(\xi)\equiv(2\pi\sigma^2)^{-1/2}\exp(-\xi^2/(2\sigma))$ is the single-variable Gaussian function. We claim that $\mathcal{N}_\sigma(\hat{\rho}) = \text{Tr}_{2,3}\left[\tilde{N}(\hat{\rho}\otimes|Q\rangle\langle Q|\otimes|P\rangle\langle P|)\right]$ where 
\begin{widetext}
\begin{align}
    \tilde{N}(\hat{\rho}\otimes|Q\rangle\langle Q|\otimes|P\rangle\langle P|) \equiv \text{SUM}_{1\rightarrow3}^\dagger\text{SUM}_{2\rightarrow1}(\hat{\rho}\otimes|Q\rangle\langle Q|\otimes|P\rangle\langle P|)\text{SUM}_{2\rightarrow1}^\dagger\text{SUM}_{1\rightarrow3}.
\end{align}    
As a result, the complementary channel of $\mathcal{N}_\sigma(\hat{\rho})$ is $\mathcal{N}^c_\sigma(\hat{\rho}) \equiv \text{Tr}_{1}\left[\tilde{N}(\hat{\rho}\otimes|Q\rangle\langle Q|\otimes|P\rangle\langle P|)\right]$. We prove the claim by noticing that
\begin{equation}
    \begin{aligned}
        \text{SUM}_{2\rightarrow1}(\hat{\rho}\otimes|\hat{q}_2=\xi_q\rangle\langle \hat{q}_2=\xi'_q|)\text{SUM}_{2\rightarrow1}^\dagger
        &=e^{-i\hat{q}_2\hat{p}_1}(\hat{\rho}\otimes|\hat{q}_2=\xi_q\rangle\langle \hat{q}_2=\xi'_q|)e^{i\hat{q}_2\hat{p}_1}\\
        &=\left(e^{-i\xi_q\hat{p}_1}\hat{\rho}e^{i\xi_{q}'\hat{p}_1}\right)\otimes|\hat{q}_2=\xi_q\rangle\langle \hat{q}_2=\xi'_q|,\\
        \text{SUM}_{1\rightarrow3}^\dagger(\hat{\rho}\otimes|\hat{p}_3=\xi_p\rangle\langle \hat{p}_3=\xi'_p|)\text{SUM}_{1\rightarrow3}&=e^{i\hat{q}_1\hat{p}_3}(\hat{\rho}\otimes|\hat{p}_3=\xi_p\rangle\langle \hat{p}_3=\xi'_p|)e^{-i\hat{q}_1\hat{p}_3}\\
        &=\left(e^{i\xi_p\hat{q}_1}\hat{\rho}e^{-i\xi_p'\hat{q}_1}\right)\otimes|\hat{p}_3=\xi_p\rangle\langle \hat{p}_3=\xi'_p|,
    \end{aligned}
\end{equation}
where we used the fact that $\hat{q}|\hat{q}=q\rangle=q|\hat{q}=q\rangle$ and $\hat{p}|\hat{p}=p\rangle=p|\hat{p}=p\rangle$. With that, we have
\begin{equation}
\begin{aligned}
    \text{Tr}_{2,3}\left[\tilde{N}(\hat{\rho}\otimes|Q\rangle\langle Q|\otimes|P\rangle\langle P|)\right] &=\int_{-\infty}^\infty d\xi_q\int_{-\infty}^\infty d\xi_pP_\sigma(\xi_q)P_\sigma(\xi_p)e^{i{\xi}_p\hat{q}_1}e^{-i{\xi}_q\hat{p}_1}\hat{\rho}e^{i{\xi}_q\hat{p}_1}e^{-i{\xi}_p\hat{q}_1}\\
    &=\int_{-\infty}^\infty d\boldsymbol{\xi}P_\sigma(\boldsymbol{\xi})\hat{D}(\boldsymbol{\xi})\hat{\rho}\hat{D}^\dagger(\boldsymbol{\xi}),
\end{aligned}
\end{equation}
where we used the fact that $\exp(i{\xi}_p\hat{q}_1)\exp(-i{\xi}_q\hat{p}_1)=\exp[i({\xi}_p\hat{q}_1-{\xi}_q\hat{p}_1)]\exp(i\xi_p\xi_q/2)$. This proves the claim.
\end{widetext}

\subsection{Calculation of $S(\mathcal{N}^c_\sigma(\hat{\rho}_{\bar{n}}))$}

We now calculate $S(\mathcal{N}^c_\sigma(\hat{\rho}_{\bar{n}})) = S\left(\text{Tr}_{1}\left[\hat{\rho}'_{\bar{n}}\right]\right)$ where 
\begin{align}
    \hat{\rho}'_{\bar{n}}\equiv \tilde{N}(\hat{\rho}_{\bar{n}}\otimes|Q\rangle\langle Q|\otimes|P\rangle\langle P|).
\end{align}
For that, we start with the covariance matrix of the three-mode Gaussian state $\hat{\rho}_{\bar{n}}\otimes|Q\rangle\langle Q|\otimes|P\rangle\langle P|$, which reads
\begin{align}
    \mathbf{V} = \left(\bar{n}+\frac{1}{2}\right)\mathbf{I}_2\oplus \begin{bmatrix}
        \sigma^2\\
        & \frac{\alpha_Q}{\sigma^2}
    \end{bmatrix}\oplus \begin{bmatrix}
        \frac{\alpha_P}{\sigma^2}\\
        & \sigma^2
    \end{bmatrix}.
\end{align}
Here $\alpha_Q = \alpha_P = \alpha= 1/4$ due to the fact that $|Q\rangle$ and $|P\rangle$ are pure states. Under the channel $\tilde{N}$, which is a composition of two SUM gates, the covariance matrix is transformed as
\begin{align}
    \mathbf{V}\rightarrow \tilde{\mathbf{V}}=\mathbf{S}_{\tilde{N}}\mathbf{V}\mathbf{S}_{\tilde{N}}^T,
\end{align}
where
\begin{align}
    \mathbf{S}_{\tilde{N}} 
    =\begin{bmatrix}
        1 & 0 & 0 & 0 & 0 & 0 \\
        0 & 1 & 0 & 0 & 0 & 1 \\
        0 & 0 & 1 & 0 & 0 & 0 \\
        0 & 0 & 0 & 1 & 0 & 0 \\
        -1 & 0 & 0 & 0 & 1 & 0 \\
        0 & 0 & 0 & 0 & 0 & 1 
    \end{bmatrix}
    \begin{bmatrix}
        1 & 0 & 1 & 0 & 0 & 0 \\
        0 & 1 & 0 & 0 & 0 & 0 \\
        0 & 0 & 1 & 0 & 0 & 0 \\
        0 & -1 & 0 & 1 & 0 & 0 \\
        0 & 0 & 0 & 0 & 1 & 0 \\
        0 & 0 & 0 & 0 & 0 & 1 
    \end{bmatrix}.
\end{align}
In other words, the three-mode Gaussian state $\hat{\rho}'_{\bar{n}}$ corresponds to the Wigner function
\begin{align}
    W(\mathbf{x}; \tilde{\mathbf{V}}, \tilde{\mathbf{x}}) &= \frac{\exp\left[-\frac{1}{2}(\mathbf{x}-\tilde{\mathbf{x}})^T\tilde{\mathbf{V}}^{-1}(\mathbf{x}-\tilde{\mathbf{x}})\right]}{(2\pi)^N\sqrt{\det\mathbf{V}}},
\end{align}
where $\mathbf{x}=(x_1,x_2,x_3,x_4,x_5,x_6)$.
The first moment $\tilde{\mathbf{x}}$ is not important because we are only interested in the covariance matrix of $\text{Tr}_1[\hat{\rho}'_{\bar{n}}]$, denoted as $\tilde{\mathbf{V}}_{2,3}$, which is obtained by integrating out $(x_1,x_2)$ from $W(\mathbf{x}; \tilde{\mathbf{V}}, \tilde{\mathbf{x}})$. As one can show, $\tilde{\mathbf{V}}_{2,3}$ corresponds to the block of $\tilde{\mathbf{V}}$ restricted to $(x_3,x_4,x_5,x_6)$, which reads
\begin{align}
    \tilde{\mathbf{V}}_{2,3} = \begin{bmatrix}
        \sigma^2 & 0 & -\sigma^2 & 0\\
        0 & \bar{n} + \frac{1}{2} + \frac{\alpha}{\sigma^2} & 0 & 0\\
        -\sigma^2 & 0 & \bar{n} + \frac{1}{2} + \sigma^2 + \frac{\alpha}{\sigma^2} & 0\\
        0 & 0 & 0 & \sigma^2
    \end{bmatrix}.
\end{align}
Upon diagonalizing $i\mathbf{\Omega}\tilde{\mathbf{V}}_{2,3}$, and with Eq.~\eqref{eq:S_rho_general}, we arrive at 
\begin{align}
\label{eq:entropy_complementary_gaussian_random_displacement_for_thermal_state}
    S(\mathcal{N}^c_\sigma(\hat{\rho}_{\bar{n}})) = g\left(\frac{\tilde{\sigma}+\sigma^2-1}{2}\right) + g\left(\frac{\tilde{\sigma}-\sigma^2-1}{2}\right),
\end{align}
where
\begin{align}
\label{eq:def_tilde_sigma}
    \tilde{\sigma} = \sqrt{4\alpha + \sigma^2(2+4\bar{n}+\sigma^2)}.
\end{align}

\subsection{Calculation of $\lim_{\bar{n}\rightarrow\infty} I_{c}(\mathcal{N}_\sigma, \hat{\rho}_{\bar{n}})$}
Upon combining Eq.~\eqref{eq:def_Ic}, \eqref{eq:entropy_gaussian_random_displacement_for_thermal_state} and \eqref{eq:entropy_complementary_gaussian_random_displacement_for_thermal_state}, we arrive at
\begin{align*}
    I_{c}(\mathcal{N}_\sigma, \hat{\rho}_{\bar{n}}) = g(\bar{n}+\sigma^2)-&g\left(\frac{\tilde{\sigma}-\sigma^2-1}{2}\right) \\
    -& g\left(\frac{\tilde{\sigma}+\sigma^2-1}{2}\right).
\end{align*}
Since $\lim_{\bar{n}\rightarrow\infty}g(\bar{n})=\log_2(e\bar{n})$ and $\lim_{\bar{n}\rightarrow\infty}\tilde{\sigma}=2\sigma\sqrt{\bar{n}}$, per Eq.~\eqref{eq:def_g} and \eqref{eq:def_tilde_sigma}, we have that $I_{c}(\mathcal{N}_\sigma, \hat{\rho}_{\bar{n}})$ is monotonically increasing with $\bar{n}$. As a result,
\begin{align*}
    \lim_{\bar{n}\rightarrow\infty}I_{c}(\mathcal{N}_\sigma, \hat{\rho}_{\bar{n}}) = \log_2(e\bar{n}) - 2\log_2(e\sigma\sqrt{\bar{n}}) = \log_2\left(\frac{1}{e\sigma^2}\right),
\end{align*}
as claimed in Eq.~\eqref{eq:lower_bound_1}.

\bibliography{GKP_facts.bib}

\end{document}